\documentclass[aps,prx,showpacs,preprintnumbers,twocolumn,nofootinbib,superscriptaddress]{revtex4-2}
\usepackage{geometry}                
\expandafter\let\csname equation*\endcsname\relax
\expandafter\let\csname endequation*\endcsname\relax

\usepackage{graphicx}
\usepackage{amssymb}
\usepackage{dsfont}
\usepackage{epstopdf}
\usepackage{color}
\usepackage{mathtools}
\usepackage{cancel}
\usepackage{chemformula}

\let\qty\SI
\usepackage{physics}

\definecolor{red}{rgb}{1,0,0}
\definecolor{blue}{rgb}{0,0,1}
\definecolor{black}{rgb}{0,0,0}

\newcommand{\p}{\partial}
\newcommand{\eq}[1]{\begin{align}#1\end{align}}
\newcommand{\eqs}[1]{\begin{align*}#1\end{align*}}
\newcommand{\ffrac}[2]{\mbox{$\frac{#1}{#2}$}}

\newcommand{\OO}{\mathcal{O}}

\newcommand{\C}{{\mathcal{C}}}

\usepackage{calrsfs}
\newcommand{\co}{c^{\circ}}

\newcommand{\muv}{\vec{\mu}}
\newcommand{\nuv}{\vec{\nu}}

\newcommand{\lambdav}{{\vec \lambda}}
\newcommand{\zetav}{\vec{\zeta}}

\newcommand{\ellv}{{\vec \ell}}
\renewcommand{\pv}{\vec{p}}
\newcommand{\qv}{\vec{q}}
\newcommand{\rv}{\vec{r}}

\newcommand{\kv}{\vec{k}}

\newcommand{\nv}{{\vec n}}

\newcommand{\zv}{\vec{z}}
\newcommand{\Gv}{\vec{G}}
\newcommand{\Jv}{{\vec J}}
\newcommand{\etav}{{\vec \eta}}

\newcommand{\lv}{\vec{\lambda}}

\newcommand{\PP}{\mathbb{P}}
\newcommand{\Sc}{\mathbb{S}}

\begin{document}

\title[Universal Slow Dynamics of Chemical Reaction Networks]{Universal Slow Dynamics of Chemical Reaction Networks}

\vspace{.5cm}

\author{Masanari Shimada$^{*}$\footnote{* These two authors contributed equally to this work}, Pegah Behrad$^{*}$, and Eric De Giuli}
\affiliation{Department of Physics, Toronto Metropolitan University\footnote{formerly Ryerson University}, M5B 2K3, Toronto, Canada}
\vspace{.5cm}

\begin{abstract}
Understanding the emergent behavior of chemical reaction networks (CRNs) is a fundamental aspect of biology and its origin from inanimate matter. A closed CRN monotonically tends to thermal equilibrium, but when it is opened to external reservoirs, a range of behaviors is possible, including transition to a new equilibrium state, a non-equilibrium state, or indefinite growth. This study shows that slowly driven CRNs are governed by the conserved quantities of the closed system, which are generally far fewer in number than the species. Considering both deterministic and stochastic dynamics, a universal slow dynamics equation is derived with singular perturbation methods, and is shown to be thermodynamically consistent. The slow dynamics is highly robust against microscopic details of the network, which may be unknown in practical situations. { In particular, non-equilibrium states of realistic large CRNs can be sought without knowledge of bulk reaction rates.} The framework is successfully tested against a suite of networks of increasing complexity and argued to be relevant in the treatment of open CRNs as chemical machines. 
\end{abstract}

\maketitle




\section{Introduction}

The goal of theory for complex systems is often to reduce the number of degrees of freedom from a large intractable number down to something manageable, whose dynamics can then be understood intuitively. Ideally, such a reduction should be principled, mathematically well-controlled, and lead to a description in terms of universal effective variables. This challenge is especially acute in the biosciences where dizzying complexity is the norm. We address it for chemical reaction networks (CRN), which provide the substrate for biochemistry and hence biology. 

{ Model reduction for CRNs has a long history \cite{Okino98,Radulescu12,Snowden17}. Since CRNs often contain a wide range of timescales, many reduction methods exploit this by quasi-equilibrium or quasi-steady-state approximations \cite{Schneider00}. However, existing theories employ different reductions for each particular CRN, thus not leading to any universal description. This may be sufficient for detailed analysis of a particular system, but makes cross-system analysis difficult and hinders unification of diverse phenomena. Moreover, these approaches work at the level of the rate equations, ignoring stochastic effects known to be important in biochemistry \cite{Bressloff17}. They may also fail to respect thermodynamic constraints \cite{Peng23}. 

}




Here we take a different route, grounded in hydrodynamics, a branch of condensed matter physics \cite{Chaikin00}. The starting point for hydrodynamics is the observation that molecular timescales, on the order of picoseconds, are miniscule compared to macroscopic forcing timescales.  Thus most degrees of freedom relax very rapidly to a state of local thermodynamic equilibrium. However, over macroscopic distances, forcing conditions can differ, thus leading to different local equilibria. In such a {\it hydrodynamic limit} in which forcing is slow in time and gradual in space, only a subset of degrees of freedom, dictated by symmetries and conservation laws, are important. Indeed, 
conserved quantities are precisely those whose densities need to be tracked, while the other degrees of freedom relax quickly and can be neglected \footnote{If a continuous symmetry is broken, then one also needs to track the density of its associated elastic variable, like the displacement field in an elastic solid. This phenomenon will play no role here.}. Conventional continuum theories for fluids, elastic solids, liquid crystals, and others are all of this type, differing only in the assumed symmetries \cite{Chaikin00}.

This vantage is natural for chemical reaction networks, since individual reactions conserve the number of each element, leading to a large number of conservation laws. { While previous and ongoing work on CRNs in the mathematical literature has also exploited conservation laws \cite{Ren06,Lee10,Desoeuvres22}, such approaches do not incorporate thermodynamic constraints, nor do they yield a universal description. Instead we combine notions from hydrodynamics with stochastic thermodynamics \cite{Schmiedl07,Seifert12,Van-den-Broeck13,Wachtel18,Wachtel22,Avanzini23},  building thermodynamic consistency in from the beginning. This is useful to quantify CRNs as chemical machines, as will be discussed later \cite{Wachtel22,Avanzini23}. } 


In particular we consider slowly driven well-mixed physical CRNs, and show that they are governed by conserved quantities, similar to hydrodynamic theories. We derive a universal slow dynamics equation, \eqref{slow3} below, that can be applied to generic slowly-driven CRNs.  For a CRN with $N$ species and $L$ conserved quantities, the reduced theory involves only $L$ variables, which is generally the number of elements, far fewer than the number of species. We work at the large-deviations level of the particle number distribution, thus incorporating the leading stochastic effects.

This article is organized as follows. First, we define our CRN, emphasizing the role of microscopic reversibility. Then we analyze the rate equation in the slowly-driven setting, showing that a naive perturbation expansion breaks down. This is cured by a singular perturbation theory, which leads to the slow dynamics equation. We then extend the theory to include stochastic effects, and test our theory with numerical simulations, showing its broad utility. Finally we show how the theory can be extended to initial states that are far from equilibrium. 

Species are indexed with $i,j,\ldots$ while reactions are indexed with $\alpha,\beta,\ldots$. We use vector notation whenever possible. For example, stoichiometric coefficients $p_{i\alpha}$ and $q_{i\alpha}$ are also written as $\pv_\alpha$ and $\qv_\alpha$. All contractions are explicitly indicated by dots. In CRNs, many functions appear that act component-wise on different species. We write $[ \vec{f}(\nv)\vec{g}(\nv) \cdots ]$ for the vector in species space whose components are $f_j(\nv) g_j(\nv)\cdots$. For example, $[\nv^{eq} e^{\vec{\ell}}]$ has components $n_j^{eq} e^{\ell_j}$, etc. When such expressions are considered as diagonal matrices, we double the brackets, i.e. $[[\vec{f}]]$ is the matrix with elements $\delta_{ij} f_{i}$. To sum or multiply over all species we write $\sum [\vec{f}]$ and $\prod [\vec{f}]$, respectively. We also apply this notation to component-wise vectors over reactions. 

\section{Slowly driven chemical reaction networks}

We define a physical CRN as follows. We have $N$ species $X_i$, interacting with $M$ reactions $\alpha$, split into the core and the boundary interactions. Write a general core reaction as
\eq{ \label{eq1}
\sum_i p_{i \alpha} X_i \rightleftharpoons \sum_i q_{i \alpha} X_i,
}
where $p_{i \alpha}$ and $q_{i \alpha}$ are the stoichiometric coefficients for the molecular species as reactant and product, respectively. 

The number of moles of all species are collected in a vector $\nv$. Quantum mechanics requires that if a reaction $\alpha$ occurs with rate $k^+_\alpha$, then its corresponding backward reaction must occur, with rate $k^-_\alpha$. As a condition for existence of thermal equilibrium, these rates are not independent but constrained in ratio to satisfy 
\eq{ \label{LDB1}
\frac{k^+_\alpha(\nv)}{k^-_\alpha(\nv)} = e^{-\frac{1}{RT}(\Delta G)_\alpha},
} 
where $(\Delta G)_{\alpha} = (\qv_{\alpha} - \pv_{\alpha}) \cdot \Gv(\nv)$ is the difference in molar Gibbs free energy between products and reactants. 
\eqref{LDB1} is known as local detailed balance \cite{Van-den-Broeck13,Polettini14,Rao16,Wachtel22}, or microscopic reversibility \cite{Astumian18}. Importantly, it does not require or imply that the CRN be in thermal equilibrium or close to it. 
In a physical CRN, we require that all reactions in the core of the system satisfy \eqref{LDB1}. For the boundary interactions, which force the system, we consider intake and degradation pairs 
\eq{
X_i^\C \rightleftharpoons X_i,
}
with rates $\epsilon r_i^+$ and $\epsilon r_i^-$, respectively, where $\epsilon$ is a dimensionless constant. The $\C$ decoration denotes a chemostat. We explain later how the boundary interactions can be generalized.


The CRN is slowly driven when $\epsilon \ll 1$, meaning that all reservoir interactions are slow compared to internal reactions. { This condition is quite natural. Indeed, bulk reaction rates are proportional to the volume of the system, $\Omega$, while boundary rates are proportional to the surface area, $\p \Omega$. Their ratio $L^* \equiv \Omega/\p\Omega$ is a length, on the order of the linear dimension of the system. To obtain the dimensionless parameter $\epsilon$, this must be compared with a microscopic length. For example, in the case of passive diffusion across a membrane with similar concentrations on both sides of the membrane, the microscopic length is $\delta \approx D/(k_0 d)$ where $d$ is the membrane thickness and $D$ is the diffusion constant \footnote{From Fick's law, the diffusion flux at each point is $D \p \phi/\p x \approx D \Delta \phi/d,$ where $\phi$ is concentration, and the derivative is approximated by a finite difference, with $d$ the membrane thickness. Integrating over the surface gives a factor $\p \Omega$, so that $\epsilon r \co \sim D \p \Omega \co/d$, where we assume $\Delta \phi \sim \co$. Then $\epsilon \sim (D \p \Omega \co/d)/(k_0 \co \Omega)$}. $\epsilon$ can be defined as $\epsilon = \delta/L^*$, which for typical values $d \sim 10 nm$ and $D \sim 10^{-9} m^2/s$ gives $\epsilon \sim 10^{-14} m / L^*$, which is small even for microscopic systems. This example furthermore illustrates that the slow driving condition can be avoided if systems have an anomalously large surface area (as in mitochondria), if diffusion is active (as in the sodium-potassium pump), if large concentration differences are held across the membrane, or if important bulk reactions are significantly slower than $k_0$.
 
}

As a consequence of slow driving, after an initial relaxation the system will be close to a thermal equilibrium state of the core CRN. However, over long timescales it can have a non-trivial dynamics near evolving equilibria, just as a fluid that is stirred or poured will transition through a series of near-equilibria, described in that case by the Navier-Stokes equations. Our main result is an evolution equation for the slow degrees of freedom, which, as we show, correspond to conserved quantities, in precise analogy with hydrodynamics. For a generic large CRN, the conserved quantities are the number of each element (H,C,O,..) and number of free electrons, if ions are present. 








The local detailed balance condition \eqref{LDB1} can also be derived microscopically. Indeed, modern transition rate theory \cite{Gilbert90} predicts from first principles
\eq{ \label{TST2}
 k^+_\alpha(\nv) = k e^{-\frac{1}{RT}(\delta G)_\alpha(\nv)},
}
where $(\delta G)_\alpha(\nv) = G_{A_\alpha} - \sum_i p_{i \alpha} G_i(\nv)$ is the difference in molar Gibbs free energy between the activated complex and the reactants, and $k=\Omega \co k_0$ in terms of $k_0 = 1/(2\pi \beta \hbar) \approx 6 \times 10^{12}$ Hz at 300K. Here $\Omega$ is the system volume, assumed to be dominated by the solvent, and $\co$ is standard concentration 1 mol/L. \eqref{TST2} holds for both forward and backward reactions, {\it mutatis mutandis}, so that \eqref{LDB1} is satisfied. 

For ideal dilute solutions, the free energies (chemical potentials) for each species take the form
\eq{
G_i(\nv) = \mu^{\circ}_i + RT \log (n_i/\co \Omega) = RT \log (n_i/n_i^{eq})
}
where $\mu^{\circ}_i$ is the chemical potential in standard conditions, and $n_i^{eq} = \Omega \co e^{-\mu^{\circ}_i/RT}$. The forward reaction rates are thus proportional to $\prod_i (n_i/\co\Omega)^{p_{i \alpha}}$, which is the law of mass action. The flux of reaction $\alpha$ is 
\eq{
J_\alpha(\nv) & = k^+_\alpha(\nv) - k^-_\alpha(\nv) \notag \\
& = k e^{-G_{A_\alpha}/RT} \left( \prod \left[ (\nv/\nv^{eq})^{\pv_{\alpha}} \right] - \prod \left[ (\nv/\nv^{eq})^{\qv_{\alpha}} \right] \right)
}
We can write the rates of reservoir interactions in the form
\eq{
r_i^+ = r_i z^\C_i, \quad r_i^-(n_i) = r_i n_i/\Omega,
}
where $z_i^\C$ is the molar concentration of species $i$ in its reservoir; in general this can be time-dependent. The numbers $r_i$ have units of rate per mole times volume, while the factors $z^\C_i$ and $n_i$ account for the law of mass action.

Note that to precisely distinguish energy from entropy, and hence to unambiguously identify heat flows, requires a microscopic Hamiltonian \cite{Schmiedl07}. However, by comparing \eqref{LDB1} to our rate parametrization for reservoir interactions, we can write
\eq{
\frac{r_i^+}{r_i^-(n_i)} = \frac{\epsilon r_i z_i^\C}{\epsilon r_i n_i/\Omega} = e^{-\frac{1}{RT}(G_i(\nv) - W_i)},
}
where $W_i$ is the work done by the reservoir in one intake reaction. This identifies $W_i = \mu^{\co}_i + RT \log (z_i^\C/\co)$, which is just the chemical potential of species $i$ at the reservoir concentration, as expected.

To quantify how far a CRN is from equilibrium, we measure the entropy production rate 
\eq{ \label{Sdot1}
T \dot{S} = TR\sum \left[ (\kv^+-\kv^-) \log \frac{\kv^+}{\kv^-} \right] \geq 0
}
Define the N by M stoichiometric matrix $\Sc = [ \Sc^0 \; \Sc^\C ]$ where $\Sc^0_{i \alpha} = q_{i \alpha} - p_{i \alpha}$ are the stoichiometric coefficients for the core reactions, and $\Sc^\C_{i \alpha} = +1$ when $\alpha$ corresponds to a reservoir of species $i$, and 0 otherwise. Using local detailed balance, the entropy production can be rewritten as
\eq{
T \dot{S} & = -\sum \left[ (\kv^+-\kv^-) \left( \Sc^T \cdot \vec{G}(\nv)-(\Sc^{\C})^T \cdot\vec{W} \right) \right] \notag \\
& = -\vec{G}(\nv) \cdot  \Sc \cdot (\kv^+-\kv^-) + \vec{W} \cdot (\rv^+-\rv^-) \label{Sdot2} ,
}
which will be useful below. 

Note that in our setup, each species present in a reservoir has {\it two} concentrations: its dynamical concentration in the system, $n_j/\Omega$, and its concentration in the reservoir, denoted $z_j^\C$. These only become equal, in general, when the reservoir rate $r_j \to \infty$ (although we will see other situations below where they equilibrate). Thus having a fully chemostatted species, as often considered in the literature, is a strong driving limit, since the corresponding species must be added or removed faster than any reaction rate in the system to maintain its constant concentration. Our setup more naturally respects real-world constraints. 

Our choice of unimolecular reservoirs is simply for convenience. The theory trivially extends to the case where the reservoir supplies a complex $Y_j$, say written as $\sum_i \tilde p_{ji} X_i$ in terms of species, where $\tilde p_{ji}$ are positive stoichiometric coefficients. One simply replaces $n_j/\Omega$ in the external flux for the reaction with $\prod_i  (n_i/\Omega)^{\tilde p_{ji}}$.

Besides slow forcing by reservoirs, one may also consider a subset of slow internal reactions. These will be discussed below.

\section{Deterministic analysis}

To illustrate our approach, we first consider the rate equations for our model; later we will generalize our results to include stochastic effects. The rate equations are
\eq{
\p_t \nv = \Sc \cdot \Jv(\nv) 
}
where $\Jv$ is the vector of reaction fluxes. 
Separating the reactions into core and boundary, this becomes
\eq{ \label{rr1}
\p_t \nv = \Sc^0 \cdot \Jv^{\;0}(\nv) + \epsilon \; [ \;\rv \;(\zv^\C - \nv/\Omega) ],
}
where $\Jv^{\;0}(\nv)$ is the reaction flux vector for core reactions, and where $r_j=0$ if there is no reservoir for species $j$. \eqref{rr1} suggests a perturbative solution in $\epsilon: \nv = \nv^0 + \epsilon \nv^1 + \ldots$. At leading order $\p_t \nv^0 = \Sc^0 \cdot \Jv^0(\nv^0)$, which describes a closed system. The system monotonically tends to thermal equilibrium, described by $\Jv^0(\nv^0)=0$. The general steady-state solution is
\eq{ \label{n0}
\nv^0 = [\nv^{eq} e^{\ellv}] = \Omega \co [e^{-\muv^{\circ}/RT} e^{\ellv} ]
}
where we must have $(\Sc^0)^T \cdot \ellv = 0$. Such vectors $\ellv$ are the {\it conserved quantities} of the closed CRN, called `moieties' \cite{De-Martino09}. Suppose there are $L$ independent moieties. Let $\zeta$ be an N by L matrix whose columns give a basis of conserved quantities. Then we can write
\eq{ \label{ellv}
\ellv = \zeta \cdot \etav,
}
where $\etav$ is a vector in moiety space. 

To give this a physical interpretation, let ${\vec Y}=(e^-,H,C,O,\ldots)$ be a vector of elements that appear in the CRN, and write each species as an abstract sum of elements
\eq{
X_j = \sum_e \tilde\zeta_{je} Y_e,
}
defining the {\it atomic matrix} $\tilde\zeta$. The condition for conservation of element $e$ at reaction $\alpha$ is 
\eq{
0 = \sum_j q_{j \alpha} \tilde\zeta_{je}  - \sum_j p_{j \alpha} \tilde\zeta_{je} = \big((\Sc^0)^T \cdot \tilde\zeta \big)_{\alpha e},
}
showing that $\tilde\zetav_e$ is a conserved quantity. In small CRNs, the moieties can differ from the elements. For example, if carbon and oxygen only appear in the CRN in multiples of CO, then their individual concentrations, while both conserved, are degenerate. However for large CRNs where we expect our theory to be of interest, there are often no conserved quantities besides the elements. 


Comparing \eqref{ellv} with \eqref{n0} we see that nonzero $\ellv$ is equivalent to shifting chemical potentials by
\eq{
\muv^{\circ} \to \muv^{\circ} - RT \zeta \cdot \etav. 
}
Therefore $RT \etav$ corresponds to a shift in the chemical potential of moieties. ( In our rate parametrization a full transformation would require also shifting activation energies by $G_{A_\alpha} \to G_{A_\alpha} - RT \vec{p_\alpha} \cdot \zeta \cdot \etav$.) It acts as a `tilt' in the free energy landscape, which fixes the number of each moiety as
\eq{
\vec{y} = \zeta^T \cdot \nv = \zeta^T \cdot [ \nv^{eq} e^{\zeta \cdot \etav} ] + \OO(\epsilon)
}
Although this gives $L$ equations in $L$ unknowns, they cannot be explicitly solved for $\etav$. 


At the next order we have
\eq{ \label{n11}
\p_t \nv^1 = (\Sc^0) \cdot \left.\big( \p \Jv^{\;0} / \p \nv \big) \right|_{\nv^0} \cdot \nv^1 + \left. [ \rv (\zv^\C - \nv/\Omega) ]  \right|_{\nv^0} 
}
If we multiply by $\zeta^T$ we get 
\eq{ \label{n12}
\p_t \zeta^T \cdot \nv^1 = \zeta^T \cdot [ \rv (\zv^\C - \nv^0/\Omega) ],
}
which expresses moiety balance. This equation can be directly integrated:
\eqs{
\zeta^T \cdot \nv^1(t) = \int_0^t dt' \zeta^T \cdot [ \rv(t') (\zv^\C(t') - \nv^0(t')/\Omega) ] .
}
Now $\nv^0(t)$ relaxes monotonically to some thermal equilibrium state $\nv^0(\infty)$. For simplicity assume that the reservoirs are independent of time. Then we can write
\eqs{
\zeta^T \cdot \nv^1(t) & = t \zeta^T \cdot [ \rv (\zv^\C - \nv^0(\infty)/\Omega) ] \notag\\
& + \int_0^t dt' \zeta^T \cdot [ \rv (\nv^0(\infty) - \nv^0(t'))/\Omega ] .
}
The integrand in the second term goes to zero at large time, so this term will be sublinear in $t$ at large time. Moreover, there is no opportunity for special cancellations with $\zv^\C$ in the first term, since $\nv^0$ is completely independent of the reservoirs. Thus, asymptotically, 

\eq{
\zeta^T \cdot \nv^1(t) \sim t \zeta^T \cdot [ \rv (\zv^\C - \nv^0(\infty)/\Omega) ],
}
which diverges at large $t$. After a time $t \sim 1/\epsilon$ we will have $\epsilon \nv^1 \sim \nv^0$ and the perturbation series breaks down. Thus the slow driving limit $\epsilon \to 0$ is {\it singular}. This is not specific to our boundary conditions or simplifying assumptions, but is completely generic. 

This mathematical singularity has a simple physical interpretation: when the system is open to reservoirs, elements can be exchanged with the environment. Over a long time scale $t \sim 1/\epsilon$, the relevant thermal equilibrium state can be completely different from its initial value. Fortunately, this suggests a cure to the long-time divergence: we need to consider a multiple-scales asymptotic analysis \cite{Bender99,Hinch91}. We introduce the slow time $\tau = t \epsilon$, so-named because $\tau \sim \OO(1)$ when $t \sim 1/\epsilon$, and replace the time derivative by
\eqs{
\frac{\p}{\p t} \to \frac{\p}{\p t} + \epsilon \frac{\p}{\p \tau} 
}
The leading order solution remains the same, except that the coefficients $\etav$ get promoted to functions of the slow time, capturing their evolution on long time scales. The $\OO(\epsilon)$ equation becomes
\eqs{ \label{n14}
\p_\tau \nv^0 + \p_t \nv^1 = (\Sc^0) \cdot \left.\big( \p \Jv^{\;0} / \p \nv \big) \right|_{\nv^0} \cdot \nv^1 + [ \rv (\zv^\C - \nv^0/\Omega) ]
}
Multiplying by $\zeta^T$ we have
\eq{
\zeta^T \cdot \p_\tau \nv^0 + \zeta^T \cdot \p_t \nv^1 = \zeta^T \cdot [ \rv (\zv^\C - \nv^0/\Omega) ]  
}
At this order, the divergences are cured if we impose
\eq{ \label{slow1}
\zeta^T \cdot \p_\tau \nv^0 = \zeta^T \cdot [ \rv (\zv^\C - \nv^0/\Omega) ]
}
which are the slow dynamics equations. As shown below, these same equations will remain valid for stochastic dynamics. We have $L$ equations in $L$ DOF. More explicitly, 
\eq{ \label{slow2}
 \sum_{j,e'} \zeta_{je} n_j^{eq} e^{\sum_e \zeta_{je} \eta_e} \zeta_{je'} \p_\tau \eta_{e'}  = \sum_j \zeta_{je} r_j (z_j^\C - \ffrac{n_j^{eq}}{\Omega} e^{\sum_e \zeta_{je} \eta_e})  
}
Defining the matrix $M(\etav)=\zeta^T \cdot [[ \nv^{eq}  e^{\zeta \cdot {\etav}} ]] \cdot \zeta$ and 
the external flux $\Jv^\C(\etav) = [\rv (\zv^\C - \nv^{eq} e^{\zeta \cdot \etav}/\Omega)]$ 
we can write this as
\eq{ \label{slow3}
M \cdot \p_\tau \etav = \zeta^T \cdot \Jv^\C,
} 
which is our main result. \eqref{slow3} is a strongly nonlinear system of equations governing the evolution of near-equilibrium states in a slowly driven CRN. The core CRN can be completely arbitrary, as long as it is detailed balanced and closed; open cases will be treated below. The reservoirs can be forced arbitrarily on the slow time scale; that is, $\rv$ and $\zv^\C$ can be arbitrary functions of $\tau$.  In particular one can consider discontinuous step-wise forcings if needed.

Note that $\zeta^T \cdot \nv$ is simply the number of moles of each moiety; thus from \eqref{slow1} the slow dynamics equation is simply the conservation law for moieties, which under slowly-driven conditions gives a closed set of equations. Remarkably, the number of degrees of freedom is reduced from $N$ down to $L$. Moreover, for the usual case where moeities correspond to elements, these slow DOF are not arbitrary but are easily interpretable and universal across different CRNs.

If some moeities appear in the CRN but not in any reservoirs, then their concentrations are clearly conserved at their initial values; in this case the corresponding entries of the slow dynamics equations can be immediately solved, leading to a further reduction in the number of evolving DOF. 


Eq.\eqref{slow3} is also remarkably universal in form: it does not depend on any { activation energies in the core, and steady states are also independent of core chemical potentials. } It can thus be applied to poorly characterized CRNs where only the stoichiometry, chemical potentials, and reservoir interactions are known.


The slow dynamics equation describes the dissipative dynamics through near-equilibrium states. In SI \cite{SI_hydro}, we show that at leading order the dissipation depends only on the slow dynamics, and not also on $\nv^1$ as might naively be expected. In particular it takes a simple form
\eq{
T \dot{S} & = \vec{W} \cdot (\rv^+-\rv^-) + \OO(\epsilon^2)
}
which can be evaluated on the solution of \eqref{slow3}.
 
It is important to emphasize that although the system is always close to a thermal equilibrium state of the closed CRN, and although $\nv^0$ takes precisely the form of a thermal equilibrium state, the system is nevertheless open, exchanging matter and energy with its surroundings. Therefore general results for detailed-balanced systems do not apply to $\nv^0$, once $\etav$ is allowed to vary on the slow timescale.

There is a fundamental relationship between the number of reservoirs and internal properties of the CRN \cite{Polettini14}. Here we adapt the arguments of \cite{Polettini14} to our setup, to be used later in analysis of non-equilibrium steady states. Define a {\it cycle} as a vector $\vec c$ in reaction space in the kernel of $\Sc$: $\Sc \cdot \vec c=0$. This is a combination of reaction fluxes that do not affect the concentration of any species. Explicitly this condition can be written $0 = \sum_{\alpha \in core} S^0_{j\alpha} c_\alpha + c_j \delta_{j \in \C}$, where $\delta_{j \in \C}$ is one if there is a reservoir of species $j$, and 0 otherwise. Let $C$ be the number of linearly independent cycles. A subset of cycles will be spanned by the core reactions alone, with dimension $C_0$; the remainder will necessarily involve a reservoir, giving an additional $C_\C$. 

By the rank-nullity theorem of linear algebra, we have $N-L_r=M-C$, where (in full generality) $L_r$ is the number of conserved moieties in the open system. We can write $L_r = L - L_{br}$, where $L$ is the number of conserved moieties in the closed system, and $L_{br}$ is the number of conservation laws that are broken by the reservoirs. Likewise we can write $M=M_0 + M_\C$, where $M_0$ is the number of reactions in the core, and $M_\C$ is the number of reservoirs. 

In the closed system, we have $N-L=M_0-C_0$; it follows then that \cite{Polettini14, Rao16}:
\eq{
M_\C = L_{br} + C_\C :
}
the number of reservoirs equals the number of broken conservation laws, plus the number of cycles involving the reservoirs; the latter are called {\it emergent cycles} and correspond to 
effective reactions performed by the CRN. 
Existence of an emergent cycle is a necessary condition for a CRN to reach a non-equilibrium steady state. Moreover, many biologically important NESS have only a few emergent cycles \cite{Wachtel22}.

As universal and general as Eq.\ref{slow3} is, it relies upon on the assumption of slow driving. General CRNs may be bistable: the dynamics can end up in different equilibria, depending upon initial conditions. However, model systems for this phenomenon all correspond to strongly driven CRNs, typically with many chemostats \cite{Schlogl72,Ge09,Vellela09}. Thus it is not obvious {\it a priori} whether such phenomena are captured by \eqref{slow3}. In fact, as proven in Appendix 2, \eqref{slow3} has no unstable equilibria if the kernel of $\zeta$ is trivial, which is expected when $N \gg L$. Since an unstable equilibrium must exist in between stable equilibria, this implies that the slow dynamics does not have multiple equilibria. Instead there is one equilibrium, which can vary over the $\tau$ timescale as external conditions evolve.

Finally, let us discuss the trivial extension when a subset of internal reactions are slow, of relative order $\epsilon$ compared to the bulk. One simply adds the corresponding projected reaction flux to the right-hand side of Eqs.\eqref{slow1},\eqref{slow2},\eqref{slow3}. If these reactions respect all conserved quantities, then they disappear identically at leading order. Instead if some conserved quantities are broken, then they will act in a similar way to external fluxes.




\section{Stochastic analysis}

We now extend our results to include stochastic effects; in a first reading, this section can be omitted. We begin from the Doi-Peliti path integral formulation \cite{Doi76,Peliti85}, which is an exact rewriting of the chemical master equation for the full counting statistics $\PP(\nv,t)$. For a self-contained review see \cite{De-Giuli22b}. The CRN is specified by the quasi-Hamiltonian (or Liouvillian) $H=H_0+\epsilon H_\C$ with
\eq{
    H_0(\nv,\nuv) & = \sum_\alpha k_\alpha \left\{ \qty ( e^{\nuv \cdot \qty ( \vec{q}_\alpha - \vec p_\alpha ) } - 1 ) \prod \left[ \qty(\frac{\nv}{\nv^{eq}})^{\pv_{\alpha}} \right] \right.\notag\\
&    \left.+ \qty ( e^{-\nuv \cdot \qty ( \vec{q}_\alpha - \vec p_\alpha ) } - 1 ) \prod \left[ \qty (\frac{\nv}{\nv^{eq}})^{\qv_{\alpha}} \right] \right\} \\
    H_\C(\nv,\nuv) & = \sum \left[ \rv (e^{-\nuv} - 1) \nv/\Omega + \rv (e^{\nuv}-1) \zv^\C \right] .
}
The $\nuv$ variables act as a per-species bias. As explained in SI, in the macroscopic regime where particle numbers are large the leading behavior of $\rho \equiv \log \PP$ satisfies a Hamilton-Jacobi equation \cite{Kubo73,Kitahara75,Smith20} 
\eq{ \label{HJ}
\frac{\p \rho(\nv,t)}{\p t} = H(\nv, -\nabla_{\nv} \rho(\nv,t)) .
}
\eqref{HJ} goes beyond the Gaussian approximation as it includes rare trajectories between different attractors, if they exist. In the slowly driven case, we require that $\Omega \co \epsilon \gg 1$, which ensures that the slow driving is more relevant than finite-size fluctuations from the bulk of the system.

Let $\{\vec \ell_x\}_{x=1}^L$ be a basis of the left kernel (or cokernel) of the stoichiometric matrix $\mathcal K = \operatorname{coker}\Sc$, i.e., this system has $L$ conserved quantities. We solve \eqref{HJ} under the initial condition
\eq{
    \rho (\nv, t=0) & = \sum \left[ \nv \log \lv^{(0)} - \lv^{(0)} - \nv \log \frac{\nv}{e} \right] , \\
    \lv^{(0)} & = [\nv^{eq} e^{\ellv^{(0)}} ] ,
}
where $\vec \ell^{(0)}$ is an arbitrary vector in $\mathcal K$. This is a large deviation function of a Poisson distribution in the limit $n_i\to\infty$
\eqs{
    \PP(\nv, 0)  = \prod \left[ \frac{\qty(\lv^{(0)})^{\nv}e^{-\lv^{(0)}}}{(\nv/e)^{\nv}} \right] \sim \prod \left[\frac{\qty(\lv^{(0)})^{\nv}e^{-\lv^{(0)}}}{\nv!} \right]
}
with means $\lv^{(0)}$.

Introducing the two time variables $t$ and $\tau = \epsilon t$ as in the previous subsection, \eqref{HJ} is rewritten as
\eqs{
    \qty ( \pdv{t} + \epsilon \pdv{\tau} ) \rho(\nv, t, \tau) & = H(\nv, -\nabla_{\nv} \rho(\nv,t,\tau)) }
with
\eqs{
    \rho (\nv, t=0, \tau=0) & = \sum_i \qty [ n_i \log \lambda_i^{(0)} - \lambda_i^{(0)} - n_i \log \frac{n_i}{e} ] .
}
We expand the solution of this equation as $\rho(\nv, t, \tau) = \rho_0(\nv, t, \tau) + \epsilon \rho_1(\nv, t, \tau) + \cdots$. The leading equation is
\eqs{
    \pdv{t} \rho_0(\nv, t, \tau) = H_0(\nv, -\nabla_{\nv} \rho_0(\nv,t,\tau)) ,
}
which is already solved by the initial condition. We cannot determine the $\tau$-dependence of the vector $\vec \ell(\tau) \in \mathcal K$ at this order.

\begin{figure*}
    \centering
    \includegraphics{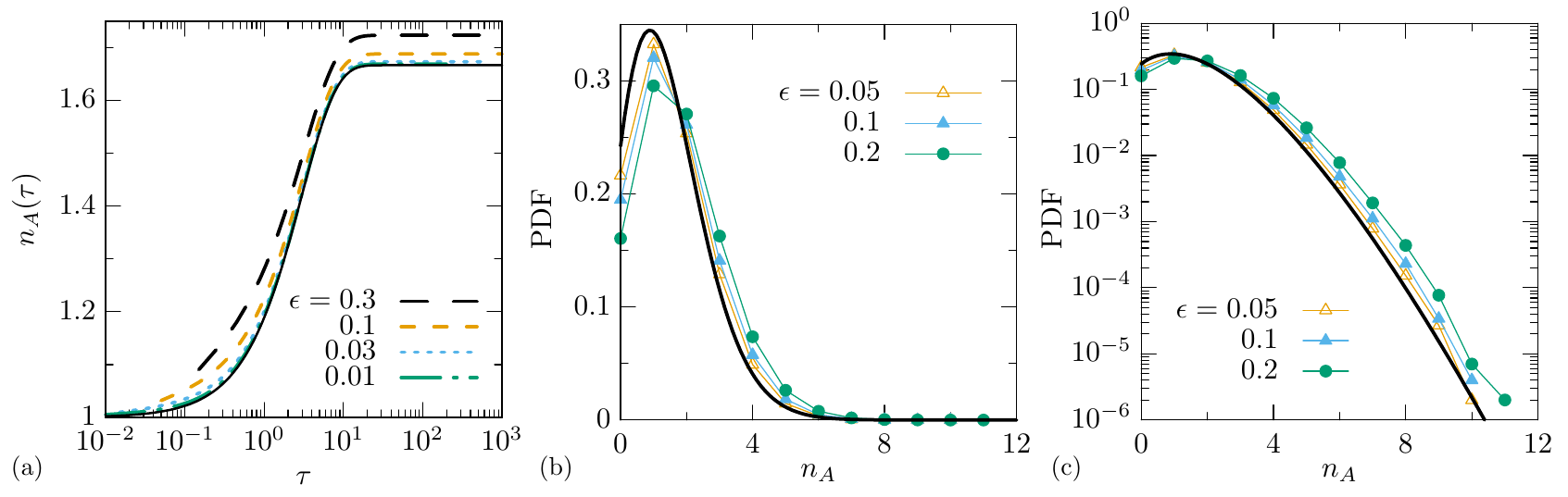}
    \caption{
    The slow dynamics equation tracks solutions to the full rate equations, even through non-equilibrium processes. In (a) the ABC model is shown with numerical solutions (dashed), and the slow dynamics equation (solid) for a range of $\epsilon$. In (b,c) the ABC model is solved at the full stochastic level, showing the distribution of species $A$ on linear and logarithmic axes, respectively. Panel (a) uses parameters $k_1 = k_2 = k_3 = r_A = r_C = 1, n_A(t=0) = n^{eq}_A = 1, n_B(t=0) = n^{eq}_B = 2, n_C(t=0) = n^{eq}_C = 3, n^\C_A = 2, n^\C_C = 4$ while panels (b,c) use parameters $r_A = 1, r_C = 1/3, n_A(t=0) = n_B(t=0) = n_C(t=0) = 0, n^{eq}_A/\Omega = 1, n^{eq}_B/\Omega = 2, n^{eq}_C/\Omega = 3, z^\C_A=2, z^\C_C=4, \Omega=3$.
    }
    \label{fig1}
\end{figure*}

\begin{figure*}
    \centering
    \includegraphics{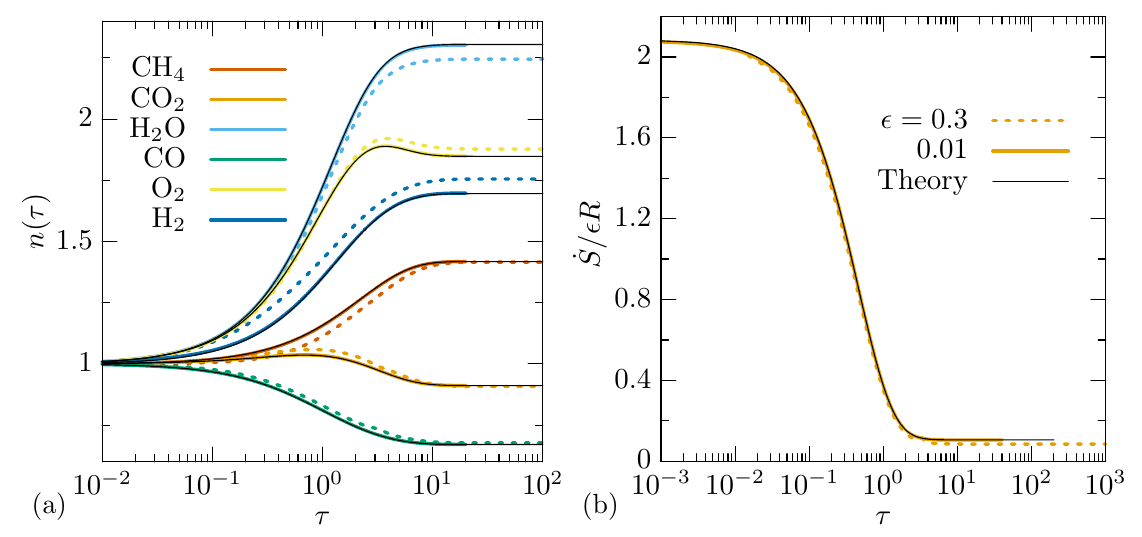}
    \caption{
    The methane combustion model is shown for (a) numerical solutions with $\epsilon=0.3$ (dashed, colored) and $\epsilon=0.01$ (solid, colored) and compared to the result from the slow dynamics equation (solid, black). From top to bottom on the right side, concentrations are shown for $\ch{H2O},\ch{O2},\ch{H2},\ch{CH4},\ch{CO2},\ch{CO}$. (b) The entropy production rate is extremely well captured by the slow dynamics, even at $\epsilon=0.3$. These results use parameters $k_i \to 1, n_i^{eq}=1, r_i = 1, n_i^\C = 2$.  
    }
    \label{fig2n}
\end{figure*}

At the next order we have
\eqs{
    & \pdv{t} \rho_1 + \pdv{\tau} \rho_0 = - \nabla_n \rho_1 \cdot \nabla_\nu H_0(\nv,-\nabla_n \rho_0) + H_\C(\nv,-\nabla_n \rho_0) ,  
    }
    which can be written
    \eq{ \label{rho1}
        \pdv{\rho_1}{t} + \nabla_n \rho_1 \cdot \nabla_\nu H_0(\nv,-\nabla_n \rho_0) = \sum \left[ (1-\nv/\lambdav) ( \p_\tau \lambdav - \Jv^{\C}) \right]
    }
    Now we note that, deterministically, the system is always close to some equilibrium state (i.e. $|\nv - \nv^{eq}| \sim \epsilon$) where $\nv^{eq}$ is of the form \eqref{n0}. Further deviations are exponentially suppressed in probability when $\Omega\co\epsilon \gg 1$, as assumed. Consider a general equilibrium $\nv = [\nv^{eq} e^{\zeta \cdot \etav'}]$ where $\etav' \neq \etav$. On any such state, we have $\nabla_\nu H_0=0$, so that this equation can be directly integrated:
  \eq{
  \rho_1|_{\nv = [\nv^{eq} e^{\zeta \cdot \etav'}]} 
  & = \int^t dt' \sum \left[ (1-e^{\zeta \cdot (\etav'-\etav)} ) ( \p_\tau \lambdav - \Jv^{\C}) \right],
  }
  which may lead to secular divergences. We demand that for all nearby equilibria $|\etav'-\etav|\ll 1$, the right-hand-side vanishes. We thus expand $e^{\zeta \cdot (\etav'-\etav)}\approx 1 + \zeta \cdot (\etav'-\etav)$ and impose 
  \eq{ \label{slowlam}
  0 = \zeta^T \cdot ( \p_\tau \lambdav - \Jv^{\C}),
  }
  which is equivalent to \eqref{slow1}, with $\nv^{(0)}$ replaced by $\vec \lambda$. We thus recover the slow dynamics equation in the stochastic approach, as the leading equation necessary to prevent long-time divergences in the singular perturbation expansion.
  
In this approximation, the particle distribution remains Poissonian at leading order, with mean $\lambdav$ that corresponds to $\nv^0$ in the rate equations. Note that the first correction to Poissonian distributions is given by the solution to \eqref{rho1}. Since this is a linear PDE for $\rho_1$, it can be solved by the method of characteristics. This solution will depend on a trajectory of the closed system.

It is clear that \eqref{slowlam} only prevents the leading divergences, and there are not enough DOF in $\lambdav(\tau)$ to prevent further ones: this implies that the full distribution must be non-Poissonian. To go beyond \eqref{slowlam}, it is easiest to use a cumulant generating function representation, as discussed in SI. This analysis shows that, once \eqref{slowlam} is solved, all higher-order divergences are tamed by solving a series of linear tensorial ODEs. These ODEs all involve the same matrix $M$ that appears in \eqref{slow3}, indicating its central role for slow dynamics in CRNs.

\section{Numerical validation}

We illustrate our theory with a series of models of increasing complexity. In realistic CRNs at room temperature, the reaction rates span a wide range of scales. This presents challenges both for numerical simulations and for the basis of our theory, which requires a time-scale separation between the bulk and the reservoir interactions. Nevertheless, at high enough temperature such a separation can be found and the theory applied. Initially we consider models for which the log-reaction rates span a modest range, corresponding to physical systems at high temperature. For such models the slow-driving condition is easily specified. Later we will consider CRNs with a wider range of rates.

\subsection{ABC Model}
The first model, which we dub the ABC model, has 3 internal reactions
\eq{
    A & \xrightleftharpoons{k_1} B , \\
    A + B & \xrightleftharpoons{k_2} 2B , \\ 
    B & \xrightleftharpoons{k_3} C
}
and two reservoirs
\eq{
    A & \xrightleftharpoons{r_A} A^\C , \\
    C & \xrightleftharpoons{r_C} C^\C .
}
It is stoichiometrically trivial, but simple enough that the slow dynamics equation can be analytically solved, as detailed in SI. This solution, giving $\nv^0(\tau)$, is compared with illustrative numerical results from the full rate equations, shown in Fig. 1a. The leading order analytical solution $\nv^0(\tau)$ differs from numerical results by an amount of order $\epsilon$, as expected.

{ The slow dynamics of the ABC model can also be solved at the stochastic level. As shown in SI, the probability remains Poissonian with computable mean. The solution agrees with direct numerical simulation of the Master equation with the Gillespie algorithm \cite{Gillespie77}, as shown in Fig 1bc. 
}

\begin{figure*}[t]
    \centering
    \includegraphics[width=\textwidth]{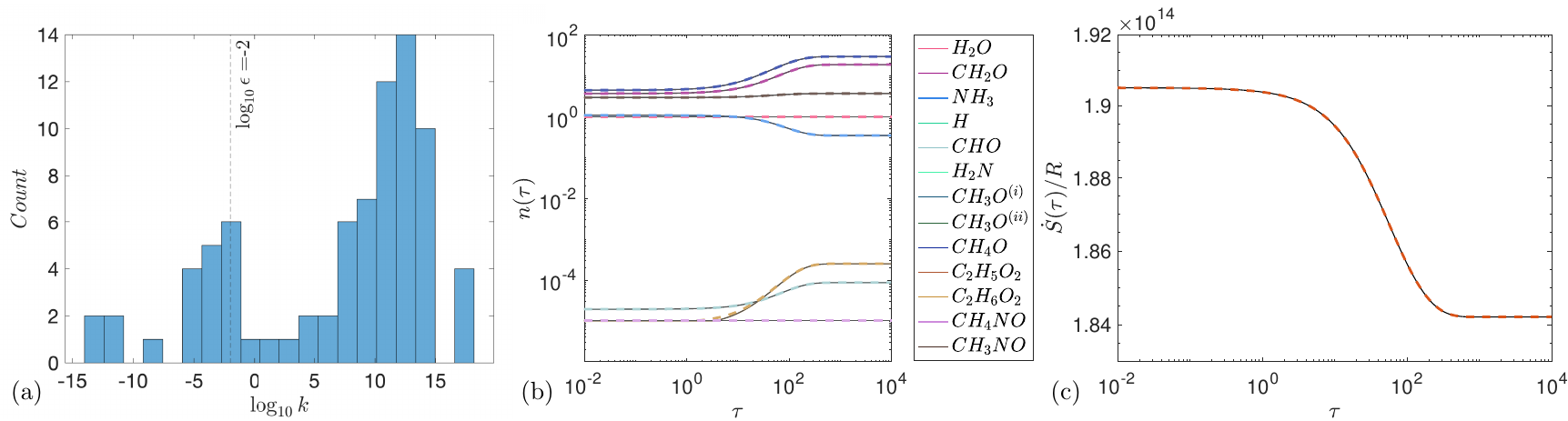}
    \caption{
    The slow dynamics equation can be applied to CRNs with a broad range of reaction rates. In (a) the histogram of reaction rates is shown (including both forward and reverse reactions), along with the chosen value of $\epsilon$ in simulations. Numerical integration of the rate equations (b, dashed, colors) is shown along with the prediction from slow dynamics (b, thin solid black). The entropy production (c, dashed colors) is well predicted by the slow dynamics contribution (c, thin solid black). 
    }
    \label{fig2}
\end{figure*}

\subsection{Methane combustion}
We now consider a version of methane combustion (see Methods):
\eq{ \label{methane}
    \ch{CH4} + \ch{3 CO2} & \xrightleftharpoons{k_1} \ch{2 H2O} + \ch{4 CO} , \\
    \ch{O2} + \ch{2 CO} & \xrightleftharpoons{k_2} \ch{2 CO2} , \\
    \ch{H2} + \ch{CO2} & \xrightleftharpoons{k_3} \ch{H2O} + \ch{CO} , \\
    \ch{H2O} & \xrightleftharpoons{r_3} \ch{H2O}^\C , \\
    \ch{O2} & \xrightleftharpoons{r_5} \ch{O2}^\C, \\
    \ch{H2} & \xrightleftharpoons{r_6} \ch{H2}^\C .
}
Although small, this CRN has features typical of large physical networks: the stoichiometric analysis (see SI) shows that there are 3 conserved quantities, corresponding to the concentrations of C, H, and O, as expected. We consider it in the high temperature limit $T \to \infty$ where all bulk rates are equal $k_i \to 1$ (in appropriate units), and we furthermore set $n_i^{eq}=1, r_i = 1, n_i^\C = 2$ for simplicity. Example numerical time evolutions are shown in Fig.2a (colored) and compared with the result from the slow dynamics equation (black). At $\epsilon=0.01$ the results are indistinguishable while even at $\epsilon=0.3$ (dashed) the slow dynamics result captures all qualitative features of the dynamics, and provides a quantitative approximation with relative errors smaller than $\epsilon$.

The entropy production rate is shown in Fig.2b (colored). As predicted by our analysis, the entropy production is well-captured by the contribution from slow dynamics (black). Thus, even though the system is always close to some thermal equilibrium state, it is nevertheless out-of-equilibrium and constantly producing entropy.

{ Moreover, for this choice of reservoirs, the final state of the system is a non-equilibrium steady state, with small but finite entropy production. It can be found by looking for steady states of the slow dynamics. This gives two equations for $e^{\eta_H}$ and $e^{\eta_O}$, which do not depend at all on the bulk dynamics. These are easily reduced to a cubic equation for $e^{\eta_O}$, with strong positivity constraints on the coefficients, since rates and concentrations cannot be negative. This situation -- reduction to a polynomial in an activity $e^{\eta_e}$ -- is typical. An explicit example will be given below. }

\subsection{Broad spectrum of reaction rates}

Here we consider an `early Earth' CRN modelling a submarine hydrothermal system containing formaldehyde, ammonia, and water, of interest for the origin of life as submarine hydrothermal systems are a potential source of abiotic amino acids \cite{Martin08,Smith16}. 
Using Reaction Mechanism Generator (see Methods) we obtain species that can be obtained from the initial pool, along with the corresponding reactions, including their rates; the full list of 13 species and 40 reactions is in SI. The histogram of reaction rates is shown in Fig. 3a: it spans 30 orders of magnitude.

This CRN conserves the concentrations of C, H, O, and N, so its slow dynamics is governed by only 4 DOF, a large reduction from the initial 13 DOF. 

We solved the rate equations with reservoirs of $\mathrm{CH_2O, H, CH_4O}$, and $\mathrm{C_{2}H_{6}O_{2}}$ at T=1000K. For this choice of reservoirs, N is still conserved, as is the difference of C and O concentrations (since all reservoirs have an equal number of C and O atoms). Thus only 2 nontrivial DOF are needed to understand its slow dynamics. As shown in Figure 3, despite the range of reaction rates spanning 30 orders of magnitude and initial concentrations spanning more than 5 orders of magnitude, the slow dynamics equation quantitatively predicts the evolution of all species, and the entropy production.

\begin{figure*}
    \centering
    \includegraphics[width=\textwidth]{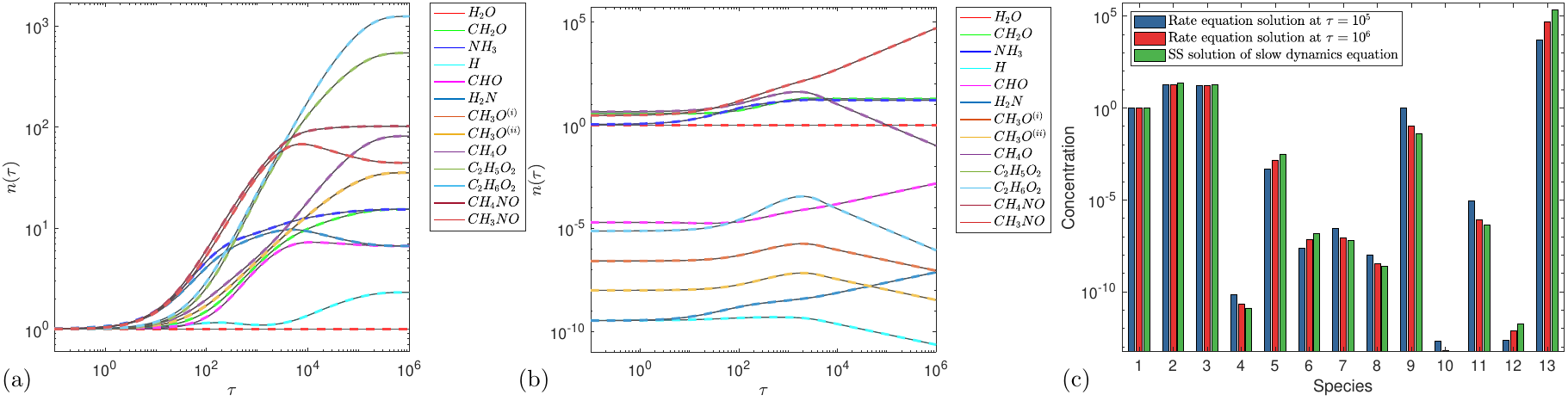}
    \caption{
    The slow dynamics equation captures extremely slow relaxations to steady states. For the `early Earth' CRN with 13 species and 40 reactions, reservoirs of $\mathrm{CH_{2}O, NH_{3}, H, CHO, H_{2}N}$ lead to very slow dynamics. In (a) the particle numbers are shown versus $\tau$ at $T=\infty$, where all bulk rates $k_\alpha=1$ in appropriate units. Here $\epsilon = 10^{-2}$. Solutions of Eq.\eqref{slow3} (thin, solid, black) track solutions to the rate equations (dashed, color), even over the 6 decades in $\tau$ needed to reach a steady state. In (b), the same CRN is simulated at $T=1000$K, where the relaxation is slower and the numerical equations very stiff. Over the simulated time, solutions to Eq.\eqref{slow3} (thin solid black) track solutions to the rate equations (dashed, color). As shown in (c), this solution appears to be converging towards a non-equilibrium steady state, found from Eq.\eqref{slow3} by looking for steady states. Species are labelled in the same order as in the legends of (a,b). For each species, three bars are shown: (from left) the solution to the rate equations after $\tau=10^5$; the solution to the rate equations after $\tau=10^6$; the steady state solution of the Eq.\eqref{slow3}.
    }
    \label{fig3}
\end{figure*}

\subsection{Long-timescale dynamics}

 The slow dynamics equation is not limited to description of mild transients between equilibrium states; the theory also applies when the concentrations are dynamic over very long time scales, or when a steady state is not reached due to varying reservoir parameters. With the early Earth CRN, if coupled to reservoirs of $\mathrm{CH_{2}O, NH_{3}, H, CHO, H_{2}N}$, then the system has a very slow dynamics. Fig.\ref{fig3} shows the dynamics both at $T=\infty$ (a) and $T=1000$K (b). In both cases, solutions of  Eq.\eqref{slow3} track solutions to the rate equations over many orders of magnitude in $\tau$. In (a), convergence to a steady state is shown; in (b), a steady state has not yet been reached, but the dynamics appears to be approaching a non-equilibrium steady state of Eq.\eqref{slow3}, as shown in Fig.\ref{fig3}c. These examples show that the final state of the CRN need not be close at all to the initial state for Eq.\eqref{slow3} to capture the dynamics.

{
\subsection{Autotrophic Core}

As a final example, we consider a very large reaction network of 404 reactions and 375 species, proposed in \cite{Wimmer21} as a minimal CRN from which to construct the amino acids and nucleic acid monomers necessary for life, along with cofactors needed for their synthesis, from primitive building blocks, namely H${}_2$, CO${}_2$, and NH${}_3$; the authors have in mind an aqueous environment like a serpentizing hydrothermal vent \cite{Smith16}. Here we show how our theory can be used with such a CRN. More details on the CRN appear in SI.


Consider first reservoirs of H${}_2$O, H${}_2$, CO${}_2$, and NH${}_3$, as suggested in \cite{Wimmer21}. These reservoirs break conservation of H, O, N, and C, and do not create any emergent cycles. Therefore the system cannot evolve to a NESS, but must relax eventually to an equilibrium state. The slow dynamics is 4 dimensional, and easily solved numerically. Since the left-hand-side of the slow dynamics equation involves the $M$ matrix, which has contributions from the equilibrium concentrations of all species, then one needs to know the chemical potentials of all bulk species. 

Consider now a scenario with more reservoirs. A minimal way to create a NESS is to add one emergent cycle. For example, if we add carbon monoxide $\ch{CO}$ then we create one emergent cycle
\eq{
\ch{CO2} + \ch{H2} & \xrightleftharpoons{} \ch{CO} + \ch{H2O}
}
Then when coupled to these reservoirs, the CRN can either grow indefinitely or evolve to a NESS, performing some of the effective reactions. NESS can be sought by looking for steady states of the slow dynamics. In terms of the net fluxes into the system from the reservoirs $J_i = r_i (z_i^\C - n_i^0/\Omega)$ these are
\eq{
0 & = J_{CO_2} + J_{CO} \\
0 & = 2 J_{H_2 O} + 2 J_{H_2} + 3 J_{NH_3} \\
0 & = 2 J_{CO_2} + J_{CO} + J_{H_2 O} \\
0 & = J_{NH_3}
}
for C, H, O, and N, respectively. The last equation means that the system must equilibrate to be at the reservoir concentration of methane. Explicitly
\eqs{
0 = r_{NH_3} (z_{NH_3}^\C - n_{NH_3}^{eq} e^{\eta_N} e^{3 \eta_H} / \Omega)
}
implies $e^{\eta_N} = e^{-3 \eta_H} z_{NH_3}^\C \Omega/n_{NH_3}^{eq}$. Note that $e^{\eta_N}$ depends on $e^{\eta_H}$, so it is still nontrivial -- but since $J_{NH_3}=0$, it drops out of the remaining equations for the NESS. 

To solve the remaining equations it is convenient to absorb equilibrium concentrations into the rates and reservoir concentrations, viz.,
\eq{
J_i = r_i (z_i^\C - n_i^{eq} e^{\sum_e \zeta_{ie} \eta_e} /\Omega) \equiv \tilde r_i (\tilde z_i - e^{\sum_e \zeta_{ie} \eta_e} )
}
Then labelling the reservoirs in the order $(\ch{H2O},\ch{H2},\ch{CO2},\ch{NH3},\ch{CO})$ the remaining equations reduce to
\eq{
0 & = \tilde r_3 (\tilde z_3 - e^{\eta_C + 2 \eta_O}) + \tilde r_5 (\tilde z_5 - e^{\eta_C + \eta_O}) \\
0 & = \tilde r_1 (\tilde z_1 - e^{2 \eta_H + \eta_O}) + \tilde r_2 (\tilde z_2 - e^{2 \eta_H }) \\
0 & = \tilde r_3 (\tilde z_3 - e^{\eta_C + 2 \eta_O}) + \tilde r_1 (\tilde z_1 - e^{2 \eta_H + \eta_O})
}
Solving the first two
\eq{
e^{\eta_C} & = e^{-\eta_O} \frac{\tilde r_3 \tilde z_3 + \tilde r_5 \tilde z_5}{\tilde r_3 e^{\eta_O} + \tilde r_5} \\
e^{2 \eta_H} & = \frac{\tilde r_1 \tilde z_1 + \tilde r_2 \tilde z_2}{\tilde r_1 e^{\eta_O} + \tilde r_2} 
}
we reduce the problem to 
\eq{ \label{CO}
0 & = \tilde r_3 \tilde z_3 + \tilde r_1 \tilde z_1 - \tilde r_3 \frac{\tilde r_3 \tilde z_3 + \tilde r_5 \tilde z_5}{\tilde r_3 e^{\eta_O} + \tilde r_5} e^{\eta_O} - \tilde r_1 \frac{\tilde r_1 \tilde z_1 + \tilde r_2 \tilde z_2}{\tilde r_1 e^{\eta_O} + \tilde r_2}  e^{\eta_O} ,
}
which becomes a quadratic equation for $e^{\eta_O}$. Thus NESS can easily be found. Now, for given reservoir rates and concentrations, one must check that each $e^{\eta_e}$ is positive; otherwise the solution is not physical. 

Physical solutions can further be divided into two classes: if all the $\eta_e$ are negative, then the equilibrium concentrations of molecules at infinite temperature decrease with increasing atomic number: larger molecules are less abundant. Instead if some $\eta_e>0$, then that element leads to an increasing concentration with increasing atomic number: large molecules can become exponentially more abundant in the NESS. This opens the possibility for a phase transition separating such regimes, which can be probed with the slow dynamics equation, with further subtleties at finite $T$ depending on the behavior of chemical potentials with atomic composition. We leave the detailed study of this effect to the future. Here, we simply solved \eqref{CO} for a variety of reservoir rates and concentrations, found solutions with all $\eta_e<0$, and then compared the result with that of the full rate equations. They agree, showing that the general structure of potential NESS can be probed without knowledge of the CRN bulk, even for very large CRNs, in the slow driving limit. 



For this CRN, we also looked at examples with numerous reservoirs. In such cases, without fine-tuning of parameters, one finds that the system grows over a range of $\tau$;  this behavior was then confirmed by solution of the full rate equations. Examples are shown in SI.

}

\section{Extension to far-from-equilibrium states}

Although above we considered initial states that were near equilibrium before coupling to external reservoirs, this is not essential to obtain a reduction to conserved quantities. Suppose instead that there is a leading order coupling to reservoirs, which we assume is stationary. Write $\rv \to \ffrac{1}{\epsilon}\rv^0 + \rv$ so that the rate equation, in the two-time {\it ansatz}, is
\eqs{
\p_t \nv + \epsilon \p_\tau \nv = \underbrace{(\Sc^0) \cdot \Jv^{\;0}(\nv) + [ \rv^0 (\nv^\C - \nv) ]}_{\tilde\Sc \cdot \tilde\Jv^{\;0}(\nv)}  + \epsilon [ \rv (\nv^\C - \nv) ]  
}
Expanding $\nv=\nv^0+\epsilon \nv^1+\ldots$, at leading order $\p_t \nv^0 = \tilde\Sc \cdot \tilde\Jv^{\;0}(\nv^0)$. With time-independent coupling to reservoirs, this will either describe relaxation to a non-equilibrium steady state (NESS), or blow-up. Assume that we reach a NESS. Then at the next order we have an equation of the form \eqref{n14}, with $\Sc$ replaced by $\tilde\Sc$. This will generally have long-time divergences unless $\nv^0$ depends on the slow time $\tau$. Let $\zeta$ be a basis of the conserved quantities of $\tilde\Sc$, i.e. $\tilde \Sc^T \cdot \zeta = 0$. Then the slow dynamics equation is again \eqref{slow1}. { The differences with the previous analysis are that now (i) the conserved quantities do not necessarily correspond to elements, since the leading-order reservoirs will break some conservation laws; and (ii) $\nv^0$ is not a known function of the slow DOF $\etav$. This latter fact means that although true, \eqref{slow1} cannot be solved without constitutive information on the $\nv^0(\etav)$ relationship. Moreover, this unknown relationship $\nv^0(\etav)$ will in general involve the bulk reaction rates, unlike the detailed-balanced case. The main result here is that one knows the dimension of the reduced dynamics, equal to the number of conserved quantities of the NESS. }

A special case that can be fully analyzed is that of an open system with no emergent cycles. In this case the system will settle to a detailed-balanced equilibrium, in which all reservoirs are equilibrated. For example, consider reservoirs of H${}_2$O, H${}_2$, CO${}_2$, and NH${}_3$. The solution \eqref{n0} will hold if the reservoir fluxes all vanish: $0=J_k$ for the four reservoir species. 
Labelling these species as $1,2,3,4$ and absorbing equilbrium concentrations into the reservoir concentrations as in Sec. VE, these become
\eq{
0 & = \tilde z_1 - e^{2\eta_H} e^{\eta_O} \\
0 & = \tilde z_2 - e^{\eta_H} \\
0 & = \tilde z_3 - e^{\eta_C} e^{2\eta_O} \\
0 & = \tilde z_4 - e^{\eta_N} e^{3\eta_H}  ,
}
which are trivially solved for the $\eta$'s. This case then reduces to that of closed detailed-balanced systems.

\section{Discussion \& Conclusions}

We have shown that slowly driven CRNs are governed by the conserved quantities of the corresponding closed system. The latter are generally the element concentrations, giving a huge reduction in DOF in large CRNs. The natural dynamical variables of the slow dynamics are chemical potentials, $\etav$, which evolve according to \eqref{slow3}. From the solution of this equation, which does not involve the bulk reaction rates, one can obtain the full dynamics, at leading order in driving. Moreover, in this limit one can easily probe the structure of NESS for large, realistic CRNs.

This framework may be useful to understand free energy transduction in open CRNs \cite{Wachtel22}. Indeed, open CRNs can be considered as chemical machines that interconvert species between the reservoirs. For example, the early Earth CRN considered above, when coupled to reservoirs of $\mathrm{CH_2O, H, CH_4O}$, and $\mathrm{C_{2}H_{6}O_{2}}$, has effective reactions (emergent cycles)
\eq{
\ch{CH_2O} + 2 \ch{H} & \xrightleftharpoons{} \ch{CH_4O} \\
\ch{CH_2O} + \ch{CH_4O} & \xrightleftharpoons{} \ch{C_{2}H_{6}O_{2}} .
}
 The free energy change in an emergent cycle is obtained straightforwardly from the chemical potentials of the species, but to understand the efficiency of the chemical machine, one requires the flux through the cycle, except in special cases \cite{Wachtel22}. Generally this necessitates the entire suite of reaction rates and a numerical solution of the rate equations. A limiting factor in the analysis is then that many rates are not known for biochemical networks of interest.

Our analysis provides an alternative. For any given forcing, one can solve the slow dynamics equation, {\it without knowledge of any bulk reaction rates}. With this solution in hand, one can evaluate the reaction fluxes and then the efficiency. This solution is guaranteed to work in the limit of slow driving, and can provide a benchmark value at finite-rate driving.

For similar reasons, our theory may be useful in conjunction with a circuit theory for CRNs \cite{Avanzini23}. In the latter, a CRN is coarse-grained by treating subsets of CRNs as chemical modules, connected to each other by particular species. For each module, the theory requires the relationship between the flux through emergent cycles and the concentrations of chemostatted species. If a module is treated as fast compared to its external connections, then our theory can be used to find the current-concentration relations, as shown in SI for an example from \cite{Avanzini23}. This is particularly useful for large, poorly-characterized modules where our method does not require the bulk reaction rates. 

How does one know when a CRN is slowly driven? First, it follows immediately from \eqref{n0} that ratios of concentrations that are stoichiometrically equivalent -- for example products and reactants of any bulk reaction -- will be constant, to leading order in $\epsilon$. This is useful if many concentrations can be tracked. More generally, one can attempt to estimate the underlying dimension of the CRN \cite{Blokhuis23}. If small, it is a strong indication that the dynamics is occurring on a slow manifold due to timescale separation.

 Independence of the slow dynamics with respect to bulk activation energies is a strong, generic form of robustness applicable to all chemical machines whose core is detailed-balanced. Moreover, dissipation is minimized, since the system is always close to some equilibrium state. The price of this robustness is that the system responds slowly to outside forcing. Whether this slow dynamics is relevant for real-world chemical machines then depends on system-specific tradeoffs between robustness and speed.

We note that an alternative weak-driving theory has been obtained in \cite{Freitas21}, also using the Hamilton-Jacobi equation. This theory is based upon the log-probability correction $\rho_1$, but without the multiple time scale analysis. This is sufficient for non-equilibrium steady states as considered in \cite{Freitas21}, but in dynamical problems it will generally suffer from long-time divergences.

Our reduction of complex dynamics to that of the conserved quantities is reminiscent of hydrodynamics. There are, however, some differences. First, in hydrodynamics one assumes that the system is coupled to other systems that differ only weakly from it. Here instead we do not assume that the reservoirs are near the system: their concentrations can be arbitrarily far from the corresponding concentration in the system. We only assume that they react slowly with the system. Second, in hydrodynamics one considers systems that interact spatially, whereas our system is well-mixed and interacts with external reservoirs without any explicit spatial coupling. The extension of our results to include spatial effects will be presented in a future publication.

%
%

\section*{Methods:}
For constructing the methane combustion model, the toolbox "Stoichiometry Tools" in MATLAB was used \cite{Kantor23}. The input $\{\ch{CH4},\ch{CO2},\ch{H2O},\ch{CO},\ch{O2},\ch{H2}\}$ led to the reactions \eqref{methane}.

For larger models, we used Reaction Mechanism Generator (RMG) \cite{Gao16,Liu21}. Given an input pool of species, RMG iteratively finds possible reactions between the species and new species that can be produced. Rates, enthalpies, and entropies are either looked up in a database, or estimated using additivity methods. For the early Earth model, we used the input set $\{\ch{CH2O},\ch{NH3},\ch{H2O}\}$, with $\ch{H2O}$ as a solvent, into RMG. It results in the 40 reactions and 10 new species shown in SI.

We note that RMG is not guaranteed to find all possible reactions among species. In testing against known CRNs relevant to the origin of life, we found that RMG sometimes failed to find reactions known to be possible. Thus we use it as a method to benchmark our framework against CRNs with valid stoichiometry and a broad range of realistic rates.

For the autotrophic core CRN, we worked only at $T \to \infty$ so that no bulk reaction rates or chemical potentials were needed. 

All ordinary differential equations were integrated in MATLAB using the \texttt{ode15s} solver. 


\section*{Acknowledgments:}
We are grateful to Mark Persic for his preliminary work on this project. This work was funded by NSERC Discovery Grant RGPIN-2020-04762 (to E. De Giuli). \\

\vfill


\bibliography{../Biology}

\end{document}


\title[Universal Slow Dynamics of Chemical Reaction Networks -- Supplementary Information]{Universal Slow Dynamics of Chemical Reaction Networks -- Supplementary Information}

\vspace{.5cm}

\author{Masanari Shimada$^{*1}$\footnote{* These two authors contributed equally to this work}, Pegah Behrad$^{*1}$, and Eric De Giuli$^1$}
\address{$^1$ Department of Physics, Toronto Metropolitan University\footnote{formerly Ryerson University}, M5B 2K3, Toronto, Canada}
\ead{edegiuli@torontomu.ca}
\vspace{.5cm}
\begin{indented}
\item[]November 2023
\end{indented}      

\vspace{1cm}

\maketitle

\section{Entropy production in the slow-driving limit}
Consider the rate of entropy production
\eq{ \label{Sdot2}
T \dot{S} & = -\vec{G}(\nv) \cdot  \Sc \cdot (\kv^+-\kv^-) + \vec{W} \cdot (\rv^+-\rv^-) 
}
The first term can be written
\eq{
-\vec{G}(\nv) \cdot  \Sc \cdot (\kv^+-\kv^-) & = -\p_t \big( \vec{G}(\nv) \cdot  \nv \big) + \nv \cdot \p_t \vec{G}(\nv) \\
& = -\p_t \big( \vec{G}(\nv) \cdot  \nv - RT \Sigma \nv \big) ,
}
which is minus the rate of change of Helmholtz free energy (the second term is pressure times volume, using the ideal gas law). Since $\vec{W} \cdot (\rv^+-\rv^-)$ is the rate of working on the system by the reservoirs, Eq. \eqref{Sdot2} is a formulation of the 1st law of thermodynamics for CRNs. 

Moreover, we can write
\eq{
T \dot{S} & = - RTk \sum_\alpha e^{-G_{A_\alpha}/RT} \left[ \sum_j S_{j\alpha} \log (n_j/n_j^{eq}) \right] \left[ \prod_i (n_i/n_i^{eq})^{p_{i \alpha}} - \prod_i (n_i/n_i^{eq})^{q_{i \alpha}} \right] + \vec{W} \cdot (\rv^+-\rv^-) \notag \\
& = - \underbrace{RTk [\log (\nv/\nv^{eq})] \cdot \Sc \cdot [[e^{-G_{\vec{A}}/RT} ]] \cdot \left[ \prod [(\nv/\nv^{eq})^{p}] - \prod [(\nv/\nv^{eq})^{q}] \right]}_{\p_t \Delta F} + \vec{W} \cdot (\rv^+-\rv^-)
}
The first term can be expanded
\eq{
\p_t \Delta F & = RTk \left[\log (\nv^0/\nv^{eq}) \left(1 + \epsilon \frac{\nv^1}{\nv^{0}} + \ldots \right) \right] \cdot \Sc \cdot [[e^{-G_{\vec{A}}/RT} ]] \cdot \notag \\
& \qquad \times \left[ \prod \left[ (\nv^0/\nv^{eq})^{p}\left(1 + p \epsilon \frac{\nv^1}{\nv^{0}} + \ldots \right) \right] - \prod \left[ (\nv^0/\nv^{eq})^{q}\left(1 + q \epsilon \frac{\nv^1}{\nv^{0}} + \ldots \right) \right] \right]
}
We have
\eq{
[\log (\nv^0/\nv^{eq})] = \vec{\ell} = \zeta \cdot \eta
}
so that $\Sc^T \cdot [\log (\nv^0/\nv^{eq})] = 0$. We also notice that
\eqs{
& \left[ \prod [ (\nv^0/\nv^{eq})^{p}\left(1 + p \epsilon \frac{\nv^1}{\nv^{0}}  \right)] \right] - \left[ \prod [ (\nv^0/\nv^{eq})^{q}\left(1 + q \epsilon \frac{\nv^1}{\nv^{0}}  \right)] \right] \notag \\
& \qquad = \left[\left[ \prod [ (\nv^0/\nv^{eq})^{p} ]\right]\right] \cdot \left(1 + \epsilon p \cdot \left[ \frac{\nv^1}{\nv^{0}} \right]  \right) - \left[\left[ \prod [ (\nv^0/\nv^{eq})^{q} ]\right]\right] \cdot \left(1 + \epsilon q \cdot \left[ \frac{\nv^1}{\nv^{0}} \right]  \right) + \ldots \\
& \qquad = \left[\left[ \prod [ (\nv^0/\nv^{eq})^{p} ] \right]\right] \cdot \epsilon p \cdot \left[ \frac{\nv^1}{\nv^{0}} \right] - \left[\left[ \prod [ (\nv^0/\nv^{eq})^{q} ]\right]\right] \cdot \epsilon q \cdot \left[ \frac{\nv^1}{\nv^{0}} \right] + \ldots
}
Combining these facts we see that $\p_t \Delta F = \OO(\epsilon^2)$. More precisely:
\eq{
\p_t \Delta F & = \epsilon^2 RTk \left[ \frac{\nv^1}{\nv^{0}} \log (\nv^0/\nv^{eq}) \right] \cdot \Sc \cdot [[e^{-G_{\vec{A}}/RT} ]] \cdot \left( \left[ \prod [ (\nv^0/\nv^{eq})^{p}] \right] \cdot  p - \left[ \prod [ (\nv^0/\nv^{eq})^{q}] \right] \cdot q  \right) \cdot \left[ \frac{\nv^1}{\nv^{0}} \right] + \ldots
}
The entropy production is written
\eq{
T \dot{S} = \epsilon \; [ \muv^{\co} + RT \log (\zv^\C/\co) ] \cdot [ \rv (\zv^\C - \nv^0/\Omega) ] + \OO(\epsilon^2),
}
and at leading order it can be determined from the slow dynamics equation, without knowledge of $\nv^1$.

\section{No unstable equilibria in slow dynamics}
\renewcommand{\mr}{\mathrm}
We consider the slow dynamics equation
\eq{
    \sum_{j,e'} \zeta_{je} \zeta_{je'} n_j^{eq} e^{\sum_e\zeta_{je}\eta_e} \pdv{\eta_{e'}}{\tau} & = - \sum_j \zeta_{je} r_j n_j^{eq} e^{\sum_e\zeta_{je}\eta_e} / \Omega + \mr{const.} 
}
Let $\eta^f_e$ be a fixed point of this equation, i.e.,
\eq{
    \sum_{j,e'} \zeta_{je} \zeta_{je'} n_j^{eq} e^{\sum_e\zeta_{je}\eta^f_e} \pdv{\eta^f_{e'}}{\tau} & = 0 .
}
Then the slow dynamics equation close to $\eta^f_e$ can be linearized as 
\eq{
    \sum_{e'} M_{ee'} \pdv{\delta \eta_{e'}}{\tau} & =  - \frac{1}{\Omega} \sum_{e'} R_{ee'} \delta \eta_{e'} + \order{\delta\eta^2} , \label{linearized equation} \\
    M_{ee'} & \coloneqq \sum_j \zeta_{je} \zeta_{je'} n_j^{eq} e^{\sum_e\zeta_{je}\eta^f_e} , \\
    R_{ee'} & \coloneqq \sum_j \zeta_{je} \zeta_{je'} r_j n_j^{eq} e^{\sum_e\zeta_{je}\eta^f_e} .
}
The matrix $\hat M$ is positive semidefinite because for an arbitrary vector $\vec x$ we have
\eq{
    \vec x \cdot \hat M \cdot \vec x & = \sum_j \qty ( \sum_e \zeta_{je} x_e )^2 n_j^{eq} e^{\sum_e\zeta_{je}\eta^f_e} > 0 .
}
Since generally $N \gg L$, we expect that the kernel of $\zeta$ is trivial, so that $M$ is positive definite. Likewise, the matrix $\hat R$ is positive semidefinite
\eq{
    \vec x \cdot \hat R \cdot \vec x & = \sum_j \qty ( \sum_e \zeta_{je} x_e )^2 r_j n_j^{eq} e^{\sum_e\zeta_{je}\eta^f_e} \geq 0 .
}
The zero modes of $R$ are vectors $\vec x$ where $\sum_e \zeta_{je} x_e=0$ for all species $j$ connected to reservoirs; these correspond to changes in $\eta$ that do not couple to reservoirs. 

As a result Eq.~\eqref{linearized equation} is rewritten as
\eq{
    \pdv{\delta \vec \eta}{\tau} & = -\frac{1}{\Omega} \hat M^{-1} \cdot \hat R \cdot \delta \vec \eta + \order{\delta \eta^2} ,\\
    \delta \vec \eta (\tau) & = \exp \qty ( - \frac{\tau}{\Omega} \hat M^{-1} \cdot \hat R ) \cdot \delta \vec \eta (0) .
}
If $\delta \vec \eta(0)$ is the eigenvector of a zero mode of $\hat R$, $\delta \vec \eta(\tau) = \delta \vec \eta(0)$. The fixed point $\vec \eta^f$ is stable in the other directions $\delta \vec \eta(\tau) \xrightarrow{\tau\to\infty} 0$. Therefore if the kernel of $\zeta$ is trivial, there are no unstable equilibria of the slow dynamics equation.

\section{Derivation of the Hamilton-Jacobi equation}

Our notation follows \cite{De-Giuli22b}. We start from the generating function
\eqs{ 
Z(\zv,t_f) = \sum_{\{\nv\}} z_1^{n_1} z_2^{n_2} \cdots z_N^{n_N} P(\nv,t_f),
}
which has a path integral representation
\eqs{ 
Z = \int \D n \int \D\nu \; e^{-S},
}
with the action $S=S_0+ S_{BC}$, where
\eq{ \label{S1}
S_0 = \int_0^{t_f} dt \left[ \nuv \cdot \p_{t} \nv - H(\nv,\nuv) \right]
}
and
\eqs{
S_{BC} = \sum_j \left[ [\nv(0)-\nv^0] \cdot \nuv(0) + n_j(t_f) [-z_j e^{-\nu_j(t_f)} + 1 - \nu_j(t_f) ] \right].
}
This gives boundary conditions $\nuv(t_f) = \log \zv$, $\nv(0) = \nv^0$. 

The particle number statistics are extracted from the generating function by contour integrals. To extract $P(\vec{m},t_f)$ we want
\eq{
P(\vec{m},t_f) & = \prod_j \oint \frac{dz_j}{2\pi i} \frac{1}{z_j^{m_j+1}} Z(\zv,t_f) \notag \\
& = \int \D n \int \D\nu \; e^{-S_0} e^{-\sum_j [\nv(0)-\nv^0] \cdot \nuv(0)} \prod_j \oint \frac{dz_j}{2\pi i} \frac{1}{z_j^{m_j+1}} e^{-n_j(t_f) [-z_j e^{-\nu_j(t_f)} + 1 - \nu_j(t_f) ]} \notag \\
& = \int \D n \int \D\nu \; e^{-S_0} e^{-\sum_j [\nv(0)-\nv^0] \cdot \nuv(0)} \prod_j \oint \frac{dz_j}{2\pi i} \frac{1}{z_j^{m_j+1}} \sum_{k_j \geq 0} \frac{1}{k_j!} \big(n_j(t_f) z_j e^{-\nu_j(t_f)} \big)^{k_j}  e^{-n_j(t_f) [1 - \nu_j(t_f) ]} \notag \\
& = \int \D n \int \D\nu \; e^{-S_0} e^{-\sum_j [\nv(0)-\nv^0] \cdot \nuv(0)} \prod_j \frac{1}{m_j!} \big(n_j(t_f) e^{-\nu_j(t_f)} \big)^{m_j}  e^{-n_j(t_f) [1 - \nu_j(t_f) ]} \notag \\
& = \int \D n \int \D\nu \; e^{-S_0} e^{-S'_{BC}}
}
with
\eq{
S'_{BC} = \sum_j \left[ [\nv(0)-\nv^0] \cdot \nuv(0) + \log (m_j!) - m_j \log n_j(t_f) + m_j \nu_j(t_f) + n_j(t_f) [1 - \nu_j(t_f) ] \right]
}
Integrating out $\nuv(t_f)$ we impose $\nv(t_f)=\vec{m}$ as expected and the other $t_f$ terms become $\log m_j! - m_j \log m_j + m_j = \OO(\log m_j)$ in the action. Then
\eq{
\rho(\vec{m},t_f) \equiv \log P(\vec{m},t_f) = \log \int_{\nv(t_f)=\vec{m}} \D [n,\nu] \; e^{-S_0} e^{-[\nv(0)-\nv^0] \cdot \nuv(0)} + (irrel)
}
The presence of a strictly fixed $\nv(0)=\nv^0$ is somewhat inconvenient. Let us regularize it by adding $\eta \nuv(0)^2$ to the action, where eventually $\eta \to 0$. Then we have
\eq{
\rho(\vec{m},0) & = \log \int d\nuv(0) e^{-[\vec{m}-\nv^0] \cdot \nuv(0)} e^{-\eta \nuv(0)^2} \notag \\
& = -[\vec{m}-\nv^0] \cdot \nuv(0) -\eta \nuv(0)^2
}
in the saddle-point approximation, giving
\eq{
\nabla_{\mv} \rho(\mv,0) = - \nuv(0).
}
Now consider a small time $t_f = \epsilon \ll 1$. In the saddle-point approximation 
\eq{
\rho(\vec{m},\epsilon) & = -\epsilon \big[ \nuv(\epsilon) \cdot \underbrace{\p_{t} \nv}_{=\frac{\p H}{\p \nuv}(\mv,\nuv(\epsilon))} - \underbrace{H(\nv(\epsilon),\nuv(\epsilon))}_{=H(\mv,\nuv(0))+(\nuv(\epsilon)-\nuv(0))\cdot \frac{\p H}{\p \nuv}(\mv,\nuv(0)) + \ldots}  \big] - [\nv(0)-\nv^0] \cdot \nuv(0) - \eta \nuv(0)^2 \notag \\
& = - \cancel{\epsilon \nuv(\epsilon) \cdot \frac{\p H}{\p \nuv}(\mv,\nuv(0))} + \epsilon H(\mv,\nuv(0))+\epsilon (\cancel{\nuv(\epsilon)}-\nuv(0))\cdot \frac{\p H}{\p \nuv}(\mv,\nuv(0)) - [\nv(0)-\nv^0] \cdot \nuv(0) - \eta \nuv(0)^2 + \ldots \notag \\
& = \epsilon H(\mv,\nuv(0)) - \nuv(0) \cdot \left[ \underbrace{\nv(0) + \epsilon \frac{\p H}{\p \nuv}(\mv,\nuv(0))}_{\nv(\epsilon) + \OO(\epsilon^2)} - \nv^0 \right]  - \eta \nuv(0)^2 \notag \\
& = \epsilon H(\mv,\nuv(0)) + \rho(\mv,0),
}
which leads to
\eq{ \label{HJSI}
\frac{\p \rho(\mv,t)}{\p t} = H(\mv, -\nabla_{\mv} \rho(\mv,t))
}
This Hamilton-Jacobi equation is Eq. 78 in \cite{Kubo73}, Eq. 2.12 in \cite{Kitahara75}, and Eq. 22 in \cite{Smith20}. The steady state version is Eq. 9 in \cite{Dykman94}.

The only approximation used in deriving Eq.\eqref{HJSI} is the saddle-point one; this includes the leading trajectories in large systems, including relaxations to equilibria and escape trajectories between different attractors. 

\section{Cumulant generating function}

Consider the cumulant generating function
\eq{
K(\vec x,t) \equiv \log \langle e^{\vec x \cdot \nv(t)} \rangle .
}
By the same derivation method as considered for the probability $P(\vec n,t)$, we can obtain a closed Hamilton-Jacobi equation for $K$, 
\eq{
    \pdv{K(\vec x, t)}{t} & = H \qty ( \pdv{K(\vec x, t)}{\vec x}, \vec x ) , \label{HJ} \\
    K(\vec x, 0) & = K_0 (\vec x) , \label{initial}
}
where the Hamiltonian is given by $H = H_0 + \epsilon H_\C$ as in the main text.
We assume that the initial state $K_0(\vec x)$ is a stationary state of $H_0$
\eq{
    0 & = H_0 \qty ( \pdv{K_0}{\vec x}, \vec x ) . \label{initial state}
}

\subsection{Moments of initial state}

The stationary distributions of $H_0$ are written as
\eq{
    \PP_0(\vec n) & = \sum_{\vec c} \rho(\vec c) \PP_{\vec c} (\vec n) ,
}
where the vector $\vec c$ denotes the conserved elements and the distribution $\PP_{\vec c}$ is a unique stationary distribution supported on the set $\{\vec n | \zeta^T\cdot \vec n = \vec c \}$, i.e., 
\eq{
    \sum_{\substack{\vec n \\ \qty ( \hat \zeta^T\cdot \vec n = \vec c ) } }\PP_{\vec c}(\vec n) = 1 .
}
Also $\rho(\vec c)$ is a distribution of the conserved elements
\eq{
    \sum_{\vec n} \PP_0(\vec n) \delta \qty ( \vec c - \hat\zeta^T \cdot \vec n ) & = \rho (\vec c) .
}
Then the cumulant generating function is given by
\eq{
    K_0 (\vec x) & = \log \sum_{\vec c} \rho(\vec c) \sum_{\vec n} e^{\vec x \cdot \vec n} \PP_{\vec c} (\vec n) \eqqcolon \log \overline{ e^{K_{\vec c}(\vec x)} }^{\vec c} ,
}
where
\eq{
    \overline{ \bullet }^{\vec c} & \coloneqq \sum_{\vec c} \rho(\vec c)\, \bullet , \\
    K_{\vec c} (\vec x) & \coloneqq \log \sum_{\vec n} e^{\vec x \cdot \vec n} \PP_{\vec c} (\vec n) .
}
If $\vec x = \hat \zeta \cdot \vec \mu\,'$ where $\mu\,'$ is an arbitrary vector with the same dimension as $\vec c$, we have
\eq{
    K_0 (\vec x = \hat \zeta\cdot \vec \mu\,') & = \log \sum_{\vec c} \rho(\vec c) \sum_{\vec n} e^{\vec \mu\,' \cdot \hat \zeta^T \cdot \vec n} \PP_{\vec c} (\vec n) = \log \sum_{\vec c} \rho(\vec c) e^{\vec \mu\,' \cdot \vec c} \eqqcolon G_0(\vec \mu\,') .
}
$G_0(\vec \mu\,')$ is the cumulant generating function of $\rho(\vec c)$.

Expanding $K_0(\vec x)$ and $G_0(\vec \mu\,')$
\eq{
    K_0(\vec x) & = \sum_{\ell=1}^\infty \frac{1}{\ell!} \sum_{j_1,\cdots,j_\ell} \kappa^{(\ell)}_{0,j_1,\cdots,j_\ell} x_{j_1} \cdots x_{j_\ell} , \\
    G_0(\vec \mu\,') = K_0 (\vec x = \hat \zeta\cdot \vec \mu\,') & = \sum_{\ell=1}^\infty \frac{1}{\ell!} \sum_{y_1,\cdots,y_\ell} \sum_{j_1,\cdots,j_\ell} \kappa^{(\ell)}_{0,j_1,\cdots,j_\ell} \zeta_{j_1,y_1} \cdots \zeta_{j_\ell,y_\ell} \mu\,'_{y_1} \cdots \mu\,'_{y_\ell} \notag \\
    & \eqqcolon \sum_{\ell=1}^\infty \frac{1}{\ell!} \sum_{y_1,\cdots,y_\ell} \gamma^{(\ell)}_{0,y_1,\cdots,y_\ell} \mu\,'_{y_1} \cdots \mu\,'_{y_\ell} ,
}
where $\kappa^{(\ell)}$ is the $\ell$-th cumulant of $\PP(\vec n)$ and $\gamma^{(\ell)}$ is that of $\rho(\vec c)$
\eq{
    \gamma^{(\ell)}_{0,y_1,\cdots,y_\ell} \coloneqq \sum_{j_1,\cdots,j_\ell} \kappa^{(\ell)}_{0,j_1,\cdots,j_\ell} \zeta_{j_1,y_1} \cdots \zeta_{j_\ell,y_\ell} .
}
For later use, we introduce
\eq{
    \left. \pdv{K_0(\vec x)}{x_j}\right|_{\vec x = \hat \zeta \cdot \vec \mu\,'} & = \left. \sum_{\ell=1}^\infty \frac{1}{(\ell-1)!} \sum_{j_1,\cdots,j_{\ell-1}} \kappa^{(\ell)}_{0,j_1,\cdots,j_{\ell-1},j} x_{j_1} \cdots x_{j_{\ell-1}} \right|_{\vec x = \hat \zeta \cdot \vec \mu\,'} \notag \\
    & = \sum_{\ell=1}^\infty \frac{1}{(\ell-1)!} \sum_{y_1,\cdots,y_{\ell-1}} \mu\,'_{y_1} \cdots \mu\,'_{y_{\ell-1}} \sum_{j_1,\cdots,j_{\ell-1}} \kappa^{(\ell)}_{0,j_1,\cdots,j_{\ell-1},j} \zeta_{j_1,y_1} \cdots \zeta_{j_{\ell-1},y_{\ell-1}} \notag \\
    & \eqqcolon \sum_{\ell=1}^\infty \frac{1}{(\ell-1)!} \sum_{y_1,\cdots,y_{\ell-1}} \mu\,'_{y_1} \cdots \mu\,'_{y_{\ell-1}} \widecheck\gamma_{0,y_1,\cdots,y_{\ell-1}; j} \eqqcolon \widecheck G_{0j} (\vec \mu\,') ,
}
where
\eq{
    \gamma^{(\ell)}_{0,y_1,\cdots,y_\ell} = \sum_j \widecheck\gamma_{0,y_1,\cdots,y_{\ell-1}; j} \zeta_{j,y_\ell} = \sum_{j_1,\cdots,j_\ell} \kappa^{(\ell)}_{0,j_1,\cdots,j_\ell} \zeta_{j_1,y_1} \cdots \zeta_{j_\ell,y_\ell} 
}

Both $\gamma^{(\ell)}_0$ and $\widecheck\gamma^{(\ell)}_0$ are determined from $\kappa_0^{(\ell)}$, but one can invert these relations. First, from the rate equation, the mean $\kappa_{0,j}^{(1)}$ is written as
\eq{
    \kappa_{0,j}^{(1)} & = n_j^{eq} e^{\sum_x \zeta_{jy} \mu_y^{(1)}} = \widecheck \gamma^{(1)}_{0,j} , \\
    \Rightarrow \gamma_{0,y}^{(1)} & = \sum_j \zeta_{jy} n_j^{eq} e^{\sum_z\zeta_{jz}\mu^{(1)}_z} .
}
Thus $\mu^{(1)}$ is a function of $\gamma^{(1)}_0$:
\eq{
    \mu^{(1)} = \mu^{(1)}(\gamma^{(1)}_0) \label{mu1}
}
and so is $\widecheck\gamma_0^{(1)}$.
Suppose that the cumulants up to the $(\ell-1)$-th order are already determined. Then from Eq.~\eqref{initial state} the equation for $\kappa^{(\ell)}$ reads
\eq{
    0 & = \sum_\alpha k_\alpha \qty { \qty ( e^{\vec S_\alpha \cdot \vec x} - 1 ) \prod_j \qty ( \frac{1}{n_j^{eq}} \pdv{K_0(\vec x)}{x_j} )^{p_{\alpha,j}} + \qty ( e^{-\vec S_\alpha \cdot \vec x} - 1 ) \prod_j \qty ( \frac{1}{n_j^{eq}} \pdv{K_0(\vec x)}{x_j} )^{q_{\alpha,j}} } \notag \\
    & = \sum_{j_1j_2\cdots j_\ell}\frac{1}{\ell!}A_{j_1\cdots j_\ell} \qty ( \kappa_0^{(1)}, \kappa_0^{(2)}, \cdots, \kappa_0^{(\ell-1)} ) x_{j_1} x_{j_2} \cdots x_{j_\ell} \notag \\
    & \qquad + \sum_\alpha k_\alpha \vec S_\alpha \cdot \vec x \prod_j \qty ( \frac{1}{n_j^{eq}} \qty ( \kappa_{0,j}^{(1)} + \frac{1}{(\ell-1)!}\sum_{j_2\cdots j_\ell}\kappa_{0,jj_2\cdots j_\ell}^{(\ell)} x_{j_2} \cdots x_{j_\ell} ) )^{p_{\alpha,j}} \notag \\
    & \qquad - \sum_\alpha k_\alpha \vec S_\alpha \cdot \vec x \prod_j \qty ( \frac{1}{n_j^{eq}} \qty ( \kappa_{0,j}^{(1)} + \frac{1}{(\ell-1)!}\sum_{j_2\cdots j_\ell}\kappa_{0,jj_2\cdots j_\ell}^{(\ell)} x_{j_2} \cdots x_{j_\ell} ) )^{q_{\alpha,j}} + \order{x^{\ell+1}} \notag \\
    & = \sum_{j_1j_2\cdots j_\ell}\frac{1}{\ell!}A_{j_1\cdots j_\ell} \qty ( \kappa_0^{(1)}, \kappa_0^{(2)}, \cdots, \kappa_0^{(\ell-1)} ) x_{j_1} x_{j_2} \cdots x_{j_\ell} \notag \\
    & \qquad + \sum_\alpha k_\alpha \vec S_\alpha \cdot \vec x \qty ( \prod_j \qty ( \frac{\kappa_{0,j}^{(1)}}{n_j^{eq}} )^{p_{\alpha,j}} ) \sum_{kj_2\cdots j_\ell} \frac{p_{\alpha,k}\kappa_{0,kj_2\cdots j_\ell}^{(\ell)}}{(\ell-1)!\kappa_{0,k}^{(1)}} x_{j_2} \cdots x_{j_\ell} \notag \\
    & \qquad - \sum_\alpha k_\alpha \vec S_\alpha \cdot \vec x \qty ( \prod_j \qty ( \frac{\kappa_{0,j}^{(1)}}{n_j^{eq}} )^{q_{\alpha,j}} ) \sum_{kj_2\cdots j_\ell} \frac{q_{\alpha,k}\kappa_{0,kj_2\cdots j_\ell}^{(\ell)}}{(\ell-1)!\kappa_{0,k}^{(1)}} x_{j_2} \cdots x_{j_\ell}  + \order{x^{j+1}} \notag \\
    & = \sum_{j_1j_2\cdots j_\ell}\frac{1}{\ell!}A_{j_1\cdots j_\ell} \qty ( \kappa_0^{(1)}, \kappa_0^{(2)}, \cdots, \kappa_0^{(\ell-1)} ) x_{j_1} x_{j_2} \cdots x_{j_\ell} \notag \\
    & \qquad + \sum_\alpha \sum_\ell S_{\alpha,\ell} x_\ell F_\alpha^+(\kappa_0^{(1)}) \sum_{kj_2\cdots j_\ell} \frac{p_{\alpha,k}\kappa_{0,kj_2\cdots j_\ell}^{(\ell)}}{(\ell-1)!\kappa_{0,k}^{(1)}} x_{j_2} \cdots x_{j_\ell} \notag \\
    & \qquad - \sum_\alpha \sum_\ell S_{\alpha,\ell} x_\ell F_\alpha^-(\kappa_0^{(1)}) \sum_{kj_2\cdots j_\ell} \frac{q_{\alpha,k}\kappa_{0,kj_2\cdots j_\ell}^{(\ell)}}{(\ell-1)!\kappa_{0,k}^{(1)}} x_{j_2} \cdots x_{j_\ell}  + \order{x^{\ell+1}} \notag \\
    & = \sum_{j_1j_2\cdots j_\ell} \frac{1}{\ell!}A_{j_1\cdots j_\ell} \qty ( \kappa_0^{(1)}, \kappa_0^{(2)}, \cdots, \kappa_0^{(\ell-1)} ) x_{j_1} x_{j_2} \cdots x_{j_\ell} \notag \\
    & \qquad - \sum_{j_1 j_2\cdots j_\ell} \frac{1}{(\ell-1)!} \sum_\alpha F_\alpha^+(\kappa_0^{(1)}) \sum_k \frac{\kappa_{0,kj_2\cdots j_\ell}^{(\ell)}S_{\alpha,k}S_{\alpha,j_1}}{\kappa_{0,k}^{(1)}} x_{j_1} x_{j_2} \cdots x_{j_\ell}  + \order{x^{\ell+1}} , \label{nth cumulant}
}
where
\eq{
    F^+_\alpha \qty (\kappa_0^{(1)}) & \coloneqq k_\alpha \prod_j \qty ( \frac{\kappa^{(1)}_{0,j}}{n_j^{eq}} )^{p_{\alpha,j}} , \\
    F^-_\alpha \qty (\kappa_0^{(1)}) & \coloneqq k_\alpha \prod_j \qty ( \frac{\kappa_{0,j}^{(1)}}{n_j^{eq}} )^{q_{\alpha,j}} = F^+_\alpha \qty ( \kappa_0^{(1)}) 
}
and the tensor $A$ is a function of $\kappa_0^{(1)}, \kappa_0^{(2)}, \cdots, \kappa_0^{(\ell-1)}$. As a result, $n$-th cumulant is generally written as
\eq{
    \kappa_{0,j_1j_2\cdots j_n}^{(\ell)} = \sum_{y_1\cdots y_\ell} \qty ( \prod_{k=1}^\ell \kappa_{0,j_k}^{(1)} \zeta_{j_ky_k} ) \mu_{y_1\cdots y_\ell}^{(\ell)} + \tilde \kappa_{0,j_1j_2\cdots j_n}^{(n)} \qty ( \mu^{(1)}, \mu^{(2)}, \cdots, \mu^{(\ell-1)} ),
}
where $\tilde \kappa^{(\ell)}_0$ satisfies Eq.~\eqref{nth cumulant} and the first term vanishes if it is contracted with the stoichiometric tensor. Note that $\tilde \kappa^{(\ell)}_0$ can depend on all the lower-order cumulants $\mu^{(\ell'\leq\ell-1)}$. The additional degrees of freedom of the tensor $\mu^{(\ell)}$ result from the element conservation. Contracting $\kappa^{(\ell)}_\ell$ with $\hat \zeta$, we have
\eq{
    \gamma^{(\ell)}_{0,y_1\cdots y_\ell} & = \sum_{z_1\cdots z_\ell} M_{y_1z_1}\cdots M_{y_\ell z_\ell} \mu^{(\ell)}_{z_1\cdots z_\ell} + \tilde \gamma^{(\ell)}_{0,y_1\cdots y_\ell} \qty ( \mu^{(1)}, \mu^{(2)}, \cdots, \mu^{(\ell-1)} ) , \\
    \Rightarrow \mu^{(\ell)}_{y_1\cdots y_\ell} & = \sum_{z_1\cdots z_\ell} M^{-1}_{y_1z_1}\cdots M^{-1}_{y_\ell z_\ell} \qty ( \gamma^{(\ell)}_{0,z_1\cdots z_\ell} - \tilde \gamma^{(\ell)}_{0,y_1\cdots y_\ell} \qty ( \mu^{(1)}, \mu^{(2)}, \cdots, \mu^{(\ell-1)} ) ) , \label{muell}
}
where
\eq{
    M_{yz} & \coloneqq \sum_j \kappa_{0j}^{(1)}\zeta_{jy}\zeta_{jz} , \\
    \tilde \gamma^{(\ell)}_{0,y_1\cdots y_\ell} & \coloneqq \sum_{j_1,\cdots,j_\ell} \tilde \kappa^{(\ell)}_{0,j_1\cdots j_\ell} \zeta_{j_1y_1} \cdots \zeta_{j_\ell y_\ell} .
}
From Eqs. \eqref{mu1} and \eqref{muell}, we see that $\mu^{(\ell)}$ is a function of $\gamma^{(1)}_0,\ldots\gamma^{(\ell-1)}_0$:
\eq{
    \mu^{(\ell)}_{y_1\cdots y_\ell} & = \sum_{z_1\cdots z_\ell} M^{-1}_{y_1z_1}\cdots M^{-1}_{y_\ell z_\ell} \qty ( \gamma^{(\ell)}_{0,z_1\cdots z_\ell} - \tilde \gamma^{(\ell)}_{0,y_1\cdots y_\ell} \qty ( \gamma_0^{(1)}, \gamma_0^{(2)}, \cdots, \gamma_0^{(\ell-1)} ) )
}
and so are $\kappa_0^{(\ell)}$ and $\widecheck\gamma_0^{(\ell)}$.

\subsection{Perturbation theory}

We first consider naive perturbation expansion
\eq{
    K(\vec x, t) & = K_0(\vec x) + \epsilon K_1(\vec x, t) + \cdots .
}
Note that the $\order{\epsilon^0}$ equation is already solved by the initial condition $K(\vec x, t=0) = K_0(\vec x)$. Then the $\order{\epsilon^1}$ equation is given by
\eq{
    \pdv{K_1(\vec x, t)}{t} = \left. \nabla_n H_0(\vec n, \vec x) \right|_{\vec n = \pdv*{K_0(\vec x)}{\vec x}} \cdot \pdv{K_1(\vec x, t)}{\vec x} + H_\C \qty ( \pdv{K_0(\vec x)}{\vec x}, \vec x ) .
}
This equation has secular divergence for $\vec x = \hat\zeta \cdot \vec \mu\,'$ as follows:
\eq{
    \pdv{K_1(\vec x = \hat\zeta \cdot \vec \mu\,', t)}{t} & = H_\C \qty ( \pdv{K_0(\vec x = \hat\zeta \cdot \vec \mu\,')}{\vec x}, \vec x = \hat\zeta \cdot \vec \mu\,' ) \notag \\
    \Rightarrow K_1(\vec x = \hat\zeta \cdot \vec \mu\,', t) & = t H_\C \qty ( \pdv{K_0(\vec x = \hat\zeta \cdot \vec \mu\,')}{\vec x}, \vec x = \hat\zeta \cdot \vec \mu\,' ) ,
}
where for illustrative purposes we assume that $H_\C$ is not explicitly time dependent. To remove this divergence we introduce another time variable $\tau = \epsilon t$ and write
\eq{
    K_0(\vec x) \xrightarrow{} K_0(\vec x, \tau) & = \sum_{\ell=1}^\infty \frac{1}{\ell!} \sum_{j_1,\cdots,j_\ell} \kappa^{(\ell)}_{j_1,\cdots,j_\ell}(\tau) x_{j_1} \cdots x_{j_\ell} .
}
Since $K_0(\vec x, \tau)$ is always a stationary solution of $H_0$, the $\tau$-dependence of $\kappa^{(\ell)}$ only results from $\rho(\vec c)$
\eq{
    \PP_0(\vec n) \xrightarrow{} \PP_0(\vec n, \tau) & = \sum_{\vec c} \rho(\vec c, \tau) \PP_{\vec c} (\vec n) .
}
Thus we only need an equation to determine the time evolution of $\rho(\vec c, \tau)$. Then we have
\eq{
    \pdv{G_0(\vec \mu\,', \tau)}{\tau} & = H_\C \qty ( \widecheck G_{0j}(\vec \mu\,', \tau) , \vec x = \hat\zeta \cdot \vec \mu\,' ) . \label{full equation}
}
Using the results in the previous section, we can solve this equation for the cumulants of $\rho(\vec c, \tau)$.


\subsection{Large deviation function}

To connect these results with those for $\rho = \log \PP$, the cumulant generating function obtained above can be decomposed as
\eq{
    K_0(\vec x, \tau) & = \tilde K (\vec x) + \delta K (\vec x, \tau) , \\
    \tilde K (\vec x) & \coloneqq \sum_j n_j^{eq} x_j + \sum_{n=2}^\infty \frac{1}{n!} \sum_{j_1\cdots j_n} \tilde \kappa_{0,j_1\cdots j_n}^{(n)} x_{j_1} \cdots x_{j_n} , \\
    \delta K (\vec x, \tau) & \coloneqq \sum_j n_j^{eq} \qty ( e^{\sum_y \zeta_{jy}\mu^{(1)}_y(\tau)} - 1 ) x_j \notag \\
    & \qquad + \sum_{n=2}^\infty \frac{1}{n!} \sum_{j_1\cdots j_n} \sum_{y_1\cdots y_n} \qty ( \prod_{k=1}^n n_{j_k}^{eq} e^{\sum_y \zeta_{j_ky}\mu_y^{(1)}(\tau)} x_{j_k} \zeta_{j_ky_k} ) \mu_{y_1\cdots y_n}^{(n)}(\tau) .
}
Note that $\delta K(\vec x, \tau=0) = 0$. Then the large deviation function of this cumulant generating function is given by
\eq{
    \rho_0( \vec n, \tau ) & = \log \int \dd \vec x\, e^{-i\vec x \cdot \vec n} e^{\tilde K (\vec x)} e^{\delta K(\vec x, \tau)} \notag \\
    & = \log \sum_{\vec n'} \tilde P(\vec n') \delta P(\vec n - \vec n', \tau) ,
}
where
\eq{
    \tilde P(\vec n) & = \int \dd \vec x\, e^{-i\vec x \cdot \vec n} e^{\tilde K (\vec x)} , \\
    \delta P(\vec n, \tau) & = \int \dd \vec x\, e^{-i\vec x \cdot \vec n} e^{\delta K (\vec x, \tau)} .
}
Note that $\delta P(\vec n, \tau=0) = \delta_{\vec n, 0}$.

Thus the approximation in the manuscript corresponds to 
\eq{
    \delta P(\vec n, \tau) \approx \delta_{\vec n, \vec\kappa^{(1)}(\tau)} ,
}
which leads to
\eq{
    \rho_0( \vec n, \tau ) & \approx \log \tilde P  [ \vec n - \vec\kappa^{(1)}(\tau) ] .
}
This is a Poisson distribution if the initial distribution is Poissonian.

\section{ABC model}
The ABC model is defined by 
\eq{
    A & \xrightleftharpoons{k_1} B , \\
    A + B & \xrightleftharpoons{k_2} 2B , \\ 
    B & \xrightleftharpoons{k_3} C
}
with two reservoirs
\eq{
    A & \xrightleftharpoons{r_A} A^\C , \\
    C & \xrightleftharpoons{r_C} C^\C . \\
}
The stoichiometric matrix for the core is 
\eq{
    \Sc^0 & = 
    \begin{pmatrix}
        \mqty{-1 \\ 1 \\ 0} & \mqty{-1 \\ 1 \\ 0} & \mqty{0 \\ -1 \\ 1}
    \end{pmatrix},
}
whose co-kernel is spanned by
\eq{
\vec \ell_1 & = \begin{pmatrix} 1 \\ 1 \\ 1 \end{pmatrix} ,
}
giving the only conserved quantity of the closed system. Then $\zeta$ is the $3 \times 1$ matrix with elements $\vec c_1$. 

The slow dynamics equation is
\eq{
(n^{eq}_A + n^{eq}_B + n^{eq}_C ) e^{\eta} \p_\tau \eta = r_A (n^\C_A - n^{eq}_A e^\eta ) + r_C (n^\C_C - n^{eq}_C e^\eta )
}
Under constant forcing this is easily solved
\eq{ \label{ABC6}
e^{\eta(\tau)} = e^{\eta(0)} e^{-a \tau} + \frac{r_A n^\C_A+r_B n^\C_B}{r_A n^{eq}_A+r_C n^{eq}_C} (1 - e^{-a \tau}),
}
where $a = (r_A n^{eq}_A+r_C n^{eq}_C )/(n^{eq}_A + n^{eq}_B + n^{eq}_C)$. 

At the stochastic level, since the ABC model has only a single conserved element, $n_{tot} = n_A + n_B + n_C$, we can write
\eq{
    \PP_{n_{tot}} (\vec n) & = \binom{n_{tot}}{n_A, n_B, n_C} p_A^{n_A} p_B^{n_B} p_C^{n_C} \delta ( n_{tot} - n_A - n_B - n_C ) \notag \\
    & = \frac{n_{tot}!}{n_A!n_B!n_C!} p_A^{n_A} p_B^{n_B} p_C^{n_C} \delta ( n_{tot} - n_A - n_B - n_C ) , \\
    \Rightarrow \PP_0 (\vec n) & = \sum_{n_{tot}} \rho(n_{tot}) \frac{n_{tot}!}{n_A!n_B!n_C!} p_A^{n_A} p_B^{n_B} p_C^{n_C} \delta ( n_{tot} - n_A - n_B - n_C ) 
}
where $p_j = n_j^{eq}/n_{tot}^{eq}$. The corresponding cumulant generating function is 
\eq{
    K_0(\vec x) & = \log \sum_{n_{tot}} \rho(n_{tot}) e^{ n_{tot} f(\vec x) } = G_0(f(\vec x)) ,
}
where
\eq{
    f(\vec x) \coloneqq \log \qty ( p_A e^{x_A} + p_B e^{x_B} + p_C e^{x_C} ) .
}
Then we have
\eq{
    \widecheck G_{j0}(x) =  \left.\pdv{K_0(\vec x)}{x_j}\right|_{\vec x = x(1,1,1)} & = \left.\pdv{G_0(f(\vec x))}{x_j} \right|_{\vec x = x(1,1,1)} = p_j \pdv{G_0(x)}{x} \label{K0G0}
}

Let us consider the following perturbation:
\eq{
H_\C = r_A \qty ( \qty ( e^{-\nu_A} - 1 ) n_A/\Omega + \qty ( e^{\nu_A} - 1 ) z_A^\C ) + r_C \qty ( \qty ( e^{-\nu_C} - 1 ) n_C/\Omega + \qty ( e^{\nu_C} - 1 ) z_C^\C ) .
}
Using the relation in Eq. \eqref{K0G0}, Eq. \eqref{full equation} simplifies to 
\eq{
    \pdv{K_0(x, \tau)}{\tau} & = r_A \qty ( e^{-x} - 1 ) (\p_{x_A} K_0)/\Omega + r_A \qty ( e^{x} - 1 ) z_A^\C  + r_C  \qty ( e^{-x} - 1 ) (\p_{x_C} K_0)/\Omega + r_C \qty ( e^{x} - 1 ) z_C^\C , \\
    \Rightarrow \pdv{G_0(x, \tau)}{\tau} & = \frac{1}{\tau_0} \qty ( e^{-x} - 1 ) \p_{x} G_0(x, \tau) + r_{tot}^\C \qty ( e^{x} - 1 ) 
}
with
\eq{
1/\tau_0 & = (r_A p_A + r_C p_C)/\Omega \\
r_{tot}^\C & = r_A z_A^{\C} + r_C z_C^{\C} .
}

We can solve this with the method of characteristics. We have
\eq{
d\tau/ds & = 1 \\
dx/ds & = - \frac{1}{\tau_0} ( e^{-x} - 1 ) \\
dG/ds & =  r_{tot}^\C (e^x-1)
}
giving
\eq{
\frac{d}{ds} \log (e^x-1) = \frac{e^{x} dx/ds}{e^x-1} = \frac{1}{\tau_0}
}
and then
\eq{
\log (e^{x(s)}-1) = \log (e^{x(0)}-1) + s / \tau_0.
}
Thus 
\eq{
dG/ds = r_{tot}^\C (e^{x(0)}-1) e^{s / \tau_0}
}
so that
\eq{
G(s) & = G(0) + (e^{x(0)}-1) r_{tot}^\C \tau_0  \qty ( e^{s /\tau_0 } - 1 ) \notag \\
& = G(x(0),0) + r_{tot}^\C \tau_0  \qty ( 1 - e^{ - s /\tau_0 } ) ( e^{x(s)} - 1 )  .
}
For simplicity, we assume that the system has no particles at first $\PP_0(\vec n,0) = \delta(\vec n)$, or $G(\vec x, 0) = 0$. As a result we have
\eq{
    G(x,\tau) & = r_{tot}^\C \tau_0  \qty ( 1 - e^{ - \tau /\tau_0 } ) ( e^x - 1 ) .
}
This is the cumulant generating function of the Poisson distribution with mean $\lambda(\tau) \coloneqq r_{tot}^\C \tau_0 \qty ( 1 - e^{ - \tau /\tau_0 } )$ and thus
\eq{
    \rho (n_{tot}, \tau) & = \frac{e^{-\lambda(\tau)}\lambda(\tau)^{n_{tot}}}{n_{tot}!} ,
}
which leads to
\eq{
    \PP_0(\vec n, \tau) & = \sum_{n_{tot}} \frac{e^{-\lambda(\tau)}\lambda(\tau)^{n_{tot}}}{n_{tot}!} \frac{n_{tot}!}{n_A!n_B!n_C!} p_A^{n_A} p_B^{n_B} p_C^{n_C} \delta ( n_{tot} - n_A - n_B - n_C ) \notag \\
    & = \frac{e^{-\sum_{j=A,B,C}p_j\lambda(\tau)}}{n_A!n_B!n_C!} \prod_{j=A,B,C} ( p_j\lambda(\tau) )^{n_j} \notag \\
    & = \prod_{j=A,B,C} \frac{e^{-p_j\lambda(\tau)}\qty( p_j\lambda(\tau) )^{n_j}}{n_j!} , \label{analytic}
}
where we used $p_A+p_B+p_C=1$. Therefore, starting from $P_0(\vec n,0) = \delta(\vec n)$, the number of each species independently follows a Poisson distribution with mean 
\eq{
    p_j\lambda(\tau) & = n_j^{eq} \frac{r_A n_A^\C + r_C n_C^\C}{r_An_A^{eq} + r_Cn_C^{eq}} \qty ( 1 - e^{ - \tau \qty ( r_A p_A + r_C p_C ) } ) .
}

\subsection{Numerical demonstration}
\begin{figure}
    \centering
    \includegraphics{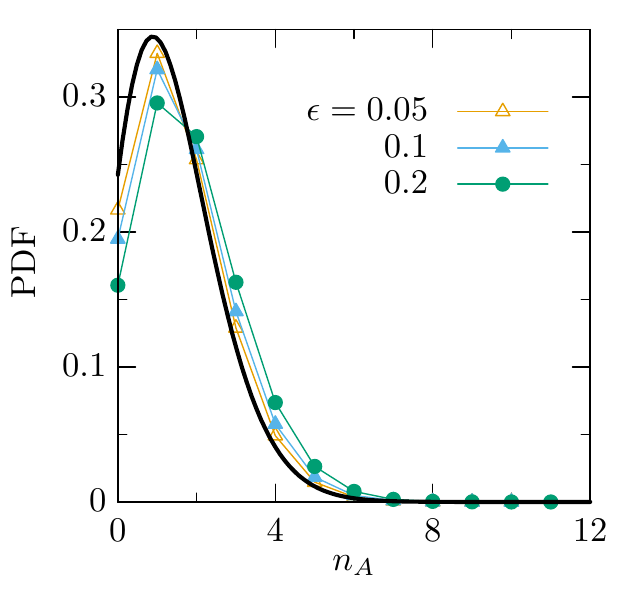}
    \caption{
    The distribution of $n_A$ at $\tau=1$. The analytical solution in Eq. \eqref{analytic} is shown by the solid line. To plot these distributions, we generated $10^6$ trajectories for each $\epsilon$.
    }
    \label{fig:abc_histogram_from_zero}
\end{figure}

We solved the full master equation of the ABC model with the following parameters:
\eq{
    n_A^{eq}/\Omega & = 1 , \\
    n_B^{eq}/\Omega & = 2 , \\
    n_C^{eq}/\Omega & = 3 , \\
    z_A^\C & = 2 , \\
    z_C^\C & = 4 , \\
    r_A & = 1 , \\
    r_C & = 1/3 , \\
    \Omega & = 3
}
and the initial condition used in the previous section
\eq{
    \PP(\vec n, t=0) & = \delta ( \vec n ) . \label{initial numerical}
}
Using Gillespie's direct method \cite{Gillespie77}, we generated trajectories consistent with the master equation. In Fig. \ref{fig:abc_histogram_from_zero}, we plotted the distributions of $n_A$ at $\tau=1$ for several values of $\epsilon$. The analytical solution in Eq. \eqref{analytic} is shown by the solid line. To plot these distributions, we generated $10^6$ trajectories for each $\epsilon$.

\begin{figure}
    \centering
    \includegraphics{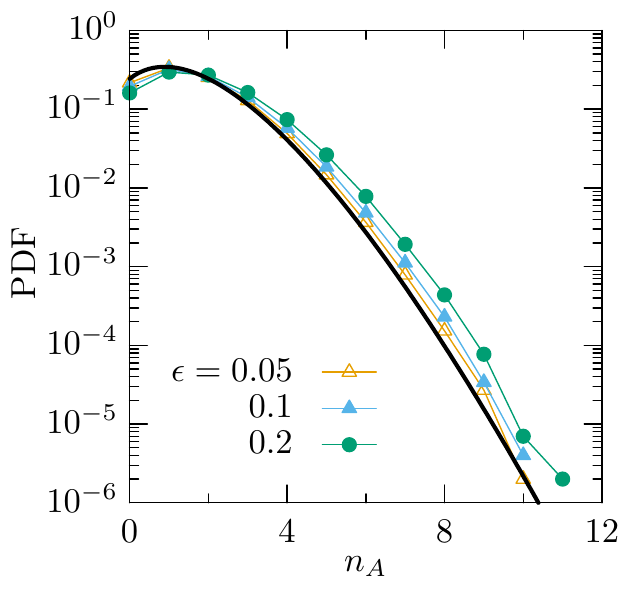}
    \caption{
    The semi-log plot of the same data as in Fig.~\ref{fig:abc_histogram_from_zero}.
    }
    \label{fig:abc_histogram_from_zero_log}
\end{figure}

\section{Methane combustion}
We now consider a version of methane combustion:
\eq{
    \ch{CH4} + \ch{3 CO2} & \xrightleftharpoons{k_1} \ch{2 H2O} + \ch{4 CO} , \\
    \ch{O2} + \ch{2 CO} & \xrightleftharpoons{k_2} \ch{2 CO2} , \\
    \ch{H2} + \ch{CO2} & \xrightleftharpoons{k_3} \ch{H2O} + \ch{CO} , \\
    \ch{H2O} & \xrightleftharpoons{r_3} \ch{H2O}^\C , \\
    \ch{O2} & \xrightleftharpoons{r_5} \ch{O2}^\C, \\
    \ch{H2} & \xrightleftharpoons{r_6} \ch{H2}^\C .
}
We label these chemical species as follows:
\eq{
    1: &\ \ch{CH4}, \\
    2: &\ \ch{CO2}, \\
    3: &\ \ch{H2O}, \\
    4: &\ \ch{CO} , \\
    5: &\ \ch{O2} , \\
    6: &\ \ch{H2} .
}
The stoichiometric matrix is
\eq{
    \Sc^0 & = 
    \begin{pmatrix}
        \mqty{-1 \\ -3 \\ 2 \\ 4 \\ 0 \\ 0} & \mqty{0 \\ 2 \\ 0 \\ -2 \\ -1 \\ 0} & \mqty{0 \\ -1 \\ 1 \\ 1 \\ 0 \\ -1}        
    \end{pmatrix}
}
with left kernel 
\eq{
    \mathcal K & = \operatorname{span} \qty { \vec \ell_1, \vec \ell_2, \vec \ell_3 } ,
}
where
\eq{
    \vec \ell_1 = \begin{pmatrix} 1 \\ 1 \\ 0 \\ 1 \\ 0 \\ 0 \end{pmatrix} ,\ 
    \vec \ell_2 = \begin{pmatrix} 4 \\ 0 \\ 2 \\ 0 \\ 0 \\ 2 \end{pmatrix} ,\ 
    \vec \ell_3 = \begin{pmatrix} 0 \\ 2 \\ 1 \\ 1 \\ 2 \\ 0 \end{pmatrix} .
}
These vectors correspond to the numbers of $\ch{C}$, $\ch{H}$, and $\ch{O}$, respectively. 

\section{Early Earth}
We consider a simple model of the early Earth, with 13 species
\eq{
\mathrm{H_{2}O}, \mathrm{CH_{2}O}, \mathrm{NH_{3}}, \mathrm{H}, \mathrm{CHO}, \mathrm{H_{2}N}, \mathrm{CH_{3}O}^{\left(i\right)}, \mathrm{CH_{3}O}^{\left(ii\right)},  \\
\mathrm{CH_{4}O}, \mathrm{C_{2}H_{5}O_{2}}, \mathrm{C_{2}H_{6}O_{2}}, \mathrm{CH_{4}NO}, \mathrm{CH_{3}NO}  ,
}
(including 2 isomers of $\ch{CH_3O}$), and 40 reactions: 
\eqs{
\mathrm{CHO}+\mathrm{CH_{3}O}^{\left(i\right)} & \rightleftharpoons 2\,\mathrm{CH_{2}O}, \qquad & \mathrm{CH_{2}O}+\mathrm{CH_{3}O}^{\left(i\right)}  \rightleftharpoons \mathrm{C_{2}H_{5}O_{2}}\\ 
2\,\mathrm{CH_{3}O}^{\left(i\right)} & \rightleftharpoons \mathrm{C_{2}H_{6}O_{2}}, \qquad &\mathrm{CH_{2}O}+\mathrm{C_{2}H_{5}O_{2}} \rightleftharpoons \mathrm{CHO}+\mathrm{C_{2}H_{6}O_{2}}\\ 
\mathrm{C_{2}H_{5}O_{2}}+\mathrm{CH_{3}O}^{\left(i\right)} & \rightleftharpoons \mathrm{CH_{2}O}+\mathrm{C_{2}H_{6}O_{2}}, \qquad & \mathrm{H_{2}N}+\mathrm{CH_{3}O}^{\left(i\right)} \rightleftharpoons \mathrm{CH_{2}O}+\mathrm{NH_{3}}\\ 
\mathrm{CH_{2}O}+\mathrm{H_{2}N} & \rightleftharpoons \mathrm{CHO}+\mathrm{NH_{3}}, \qquad &\mathrm{C_{2}H_{6}O_{2}}+\mathrm{H_{2}N} \rightleftharpoons \mathrm{C_{2}H_{5}O_{2}}+\mathrm{NH_{3}}\\ 
\mathrm{CH_{2}O}+\mathrm{H_{2}N} & \rightleftharpoons \mathrm{CH_{4}NO}, \qquad & \mathrm{CHO}+H \rightleftharpoons \mathrm{CH_{2}O}, \\ 
\mathrm{CH_{2}O}+H & \rightleftharpoons \mathrm{CH_{3}O}^{\left(i\right)}, \qquad &\mathrm{C_{2}H_{5}O_{2}}+H  \rightleftharpoons \mathrm{C_{2}H_{6}O_{2}}\\ 
\mathrm{CHO}+\mathrm{H_{2}N} & \rightleftharpoons \mathrm{CH_{3}NO}, \qquad &\mathrm{CH_{3}NO}+H  \rightleftharpoons \mathrm{CH_{4}NO}\\  
\mathrm{CHO}+\mathrm{CH_{4}NO} & \rightleftharpoons \mathrm{CH_{2}O}+\mathrm{CH_{3}NO}, \qquad  &\mathrm{CH_{4}NO}+\mathrm{C_{2}H_{5}O_{2}} \rightleftharpoons \mathrm{CH_{3}NO}+\mathrm{C_{2}H_{6}O_{2}}, \\
\mathrm{CHO}+\mathrm{CH_{3}O}^{\left(ii\right)} & \rightleftharpoons 2\,\mathrm{CH_{2}O}, \qquad &\mathrm{H_{2}N}+\mathrm{CH_{3}O}^{\left(ii\right)}  \rightleftharpoons \mathrm{CH_{2}O}+\mathrm{NH_{3}}\\ 
\mathrm{CH_{3}O}^{\left(i\right)} & \rightleftharpoons \mathrm{CH_{3}O}^{\left(ii\right)}, \qquad &\mathrm{C_{2}H_{5}O_{2}}+\mathrm{CH_{3}O}^{\left(ii\right)} \rightleftharpoons \mathrm{CH_{2}O}+\mathrm{C_{2}H_{6}O_{2}}\\ 
\mathrm{CH_{2}O}+H & \rightleftharpoons \mathrm{CH_{3}O}^{\left(ii\right)}, \qquad & \mathrm{CH_{2}O}+\mathrm{CH_{3}O}^{\left(i\right)}  \rightleftharpoons \mathrm{CHO}+\mathrm{CH_{4}O}\\
\mathrm{CH_{4}O}+\mathrm{H_{2}N} & \rightleftharpoons \mathrm{NH_{3}}+\mathrm{CH_{3}O}^{\left(i\right)}, \qquad & 2\,\mathrm{CH_{3}O}^{\left(i\right)}  \rightleftharpoons \mathrm{CH_{2}O}+\mathrm{CH_{4}O}\\ 
\mathrm{CH_{4}O}+\mathrm{C_{2}H_{5}O_{2}} & \rightleftharpoons \mathrm{C_{2}H_{6}O_{2}}+\mathrm{CH_{3}O}^{\left(i\right)}, \qquad & \mathrm{CH_{4}NO}+\mathrm{CH_{3}O}^{\left(i\right)}  \rightleftharpoons \mathrm{CH_{4}O}+\mathrm{CH_{3}NO}\\ 
\mathrm{CH_{2}O}+\mathrm{CH_{3}O}^{\left(ii\right)} & \rightleftharpoons \mathrm{CHO}+\mathrm{CH_{4}O}, \qquad & \mathrm{CH_{4}O}+\mathrm{H_{2}N}  \rightleftharpoons \mathrm{NH_{3}}+\mathrm{CH_{3}O}^{\left(ii\right)}\\ 
\mathrm{CH_{3}O}^{\left(i\right)}+\mathrm{CH_{3}O}^{\left(ii\right)} & \rightleftharpoons \mathrm{CH_{2}O}+\mathrm{CH_{4}O}, \qquad & \mathrm{CH_{3}O}^{\left(i\right)}+\mathrm{CH_{3}O}^{\left(ii\right)} \rightleftharpoons \mathrm{CH_{2}O}+\mathrm{CH_{4}O}\\ 
\mathrm{CH_{4}O}+\mathrm{C_{2}H_{5}O_{2}} & \rightleftharpoons \mathrm{C_{2}H_{6}O_{2}}+\mathrm{CH_{3}O}^{\left(ii\right)}, \qquad & \mathrm{CH_{4}NO}+\mathrm{CH_{3}O}^{\left(ii\right)} \rightleftharpoons \mathrm{CH_{4}O}+\mathrm{CH_{3}NO}\\ 
H+\mathrm{CH_{3}O}^{\left(ii\right)} & \rightleftharpoons \mathrm{CH_{4}O}, \qquad & 2\,\mathrm{CH_{3}O}^{\left(ii\right)} \rightleftharpoons \mathrm{CH_{2}O}+\mathrm{CH_{4}O}\\ 
\mathrm{CH_{3}O}^{\left(i\right)}+\mathrm{CH_{3}O}^{\left(ii\right)} & \rightleftharpoons \mathrm{CH_{2}O}+\mathrm{CH_{4}O}, \qquad & \mathrm{CH_{3}O}^{\left(i\right)}+\mathrm{CH_{3}O}^{\left(ii\right)}  \rightleftharpoons \mathrm{CH_{2}O}+\mathrm{CH_{4}O}\\
\mathrm{CH_{4}O}+\mathrm{CH_{3}O}^{\left(ii\right)} & \rightleftharpoons \mathrm{CH_{4}O}+\mathrm{CH_{3}O}^{\left(i\right)}, \qquad & H+\mathrm{CH_{3}O}^{\left(i\right)}  \rightleftharpoons \mathrm{CH_{4}O} \\ 
\mathrm{CH_{4}NO}+\mathrm{H_{2}N} & \rightleftharpoons \mathrm{CH_{3}NO}+\mathrm{NH_{3}}, \qquad & H+\mathrm{H_{2}N}  \rightleftharpoons \mathrm{NH_{3}}
}



\section{Autotrophic Core} The CRN of \cite{Wimmer21} has some features, typical of biochemical reaction networks with large molecules, that require additional care. 

First, some reactions are given in an unbalanced form, omitting factors of water or hydrogen. It includes the reactions involving species like $C00028$ as a Hydrogen-acceptor and $C00030$ a Hydrogen-donor. Also, for species like $C17023$ as sulfur donor, we replace it simply with an atom of sulfur $S$. We remove from our list of reactions any remaining incomplete reactions that lack complex species and do not balance simply by the addition of simple elements like hydrogen. 

Second, some reactions are given in an incompletely specified form. For example, a reaction may involve a hydrogen-sulfide group on a biomolecule, but is agnostic to whatever complexes form the rest of the molecule. In this case the compounds are listed with `R' complexes, which balance on both sides of the reaction, and represent the unknown and largely irrelevant parts of the molecules. In the CRN of \cite{Wimmer21}, 28 of the 404 reactions involve such R groups. To treat them, we used two different methods: (i) the first method is to make each distinct R complex a new `element', and give them a concentration; (ii) the second method is to remove the R complexes entirely. These two methods are opposing limits: in the first, the imposed concentrations of each R element can rate-limit any reaction in which they occur, while in the second, by construction the R complexes never rate-limit any reaction.  

Analysing the reactions here, we found that there are seven distinct types of R complex that appear, which we labelled $R_i, i=1,\ldots,7$. The corresponding species are then modified to
\begin{eqnarray*}
&&\{\ch{C5H12O13P3R1}, \ch{C5H11O10P2R1}, \ch{HSR2}, \ch{C3H3O3SR2}, \ch{C4H5O3SR2}, \ch{C7H11O3SR2}, \ch{C8H13O3SR2}, \\
&& \ch{C6H7O4SR2}, \ch{C6H9O4SR2}, \ch{C6H7O3SR2}, \ch{C6H9O3SR2}, \ch{C8H11O4SR2}, \ch{C8H13O4SR2}, \ch{C8H11O3SR2}, \\
&& \ch{C15H21N5O10PR3}, \ch{C20H28N6O13PR3}, \ch{C5H8NO4R4}, \ch{C11H17N2O7R4}, \ch{C11H18N2O10PR4}, \\
&& \ch{C11H17N2O6R4}, \ch{C11H20N3O5R4}, \ch{C4H7N2O3R5}, \ch{C14H19N7O9PR5}, \ch{C4H7N2O2SR5}, \\
&& \ch{C6H10NO3SR6}, \ch{HSR6}, \ch{C4H6N2O2SR7}, \ch{C4H6N2O2S2R7} \}
\end{eqnarray*}

After this preprocessing, our network has 375 species and 397 reactions. The species consists of 9 elements \{C, H, N, O, P, S, Co, Mo, Ni\} and 7 types of unknown $R$ components. To construct the atomic matrix $\zeta$, the two different methods for treating $R$ complexes yield either 9 conserved quantities total, or 16. 

We solved the rate equations for several illustrative examples, using both methods of treating the R complexes. Some ionic compounds are present in the CRN. For simplicity we ignored free electrons, assuming them to be abundant.

Assuming the high temperature limit $T \to \infty$, all bulk rates are equal $k_i \to 1$ (in appropriate units). We parameterize the reservoir concentrations randomly as $\vec{z}^{\C} = \vec{n}^{eq} + 100 \exp{(-1+2 * rand()})$ and the system-reservoir interaction rate as $0.01$. Therefore, $\epsilon = 0.01$ and the simulation time is of order $1/\epsilon = 10^2$.

\subsection{Reservoirs}
In the main text, we discuss this CRN with reservoirs \{H${}_2$O, H${}_2$, CO${}_2$, and NH${}_3$\}, as suggested in \cite{Wimmer21}, and then the case with CO added. In the first case, it is clear that the reservoirs break conservation of N, C, O, and H. From the Polettini-Esposito result (see main text) it follows that no emergent cycles exist. Adding CO, we cannot break any conservation laws, so we must add one emergent cycle, which can be taken as 
\eqs{
\ch{CO2} + \ch{H2} & \xrightleftharpoons{} \ch{CO} + \ch{H2O}
}

Below, we show results for larger sets of reservoirs. We considered two sets of reservoirs. The first choice of species as reservoirs is restricted to simple compounds in the CRN, which are listed in table \ref{tab1}. The second choice uses the species listed in Table 1 of \cite{Wimmer21}, which are highly connected nodes among the autotrophic core, as reservoirs (See table \ref{tab2}). 

\begin{table}[htp]
\centering
\caption{The first choice of reservoirs} 
\begin{tabular}{c c}
\hline\hline
compound   &  KEGG   \\ 
\hline
$\ch{H2O}$ & $C00001$ \\
$\ch{CO2}$ & $C00011$  \\
$\ch{NH3}$ & $C00014$  \\
$\ch{H2}$ & $C00030$  \\
$\ch{CH2O}$ & $C00067$  \\
$\ch{H}$ & $C00080$  \\
$\ch{CO}$ & $C00237$  \\
$\ch{H2S}$ & $C00283$  \\
\hline
\end{tabular}
\label{tab1}
\caption{The first set of reservoirs chosen, along with their KEGG IDs \cite{Kanehisa00}.}
\end{table}


\begin{table}[htp]
\begin{center}
\caption{The second choice of reservoirs} 
\begin{tabular}{c c c}
\hline\hline
compound   &  KEGG   &  formula \\
\hline
Water & $C00001$  & $H_2O$  \\ 
$ATP$ & $C00002$ & $C_{10}H_{16}N_{5}O_{13}P_{3}$ \\
$NAD^+$ & $C00003$ & $C_{21}H_{28}N_{7}O_{14}P_2$ \\
$NADH$ & $C00004$ & $C_{21}H_{29}N_7O_{14}P_2$ \\
$NADPH$ & $C00005$ & $C_{21}H_{30}N_7O_{17}P_3$ \\
$NADP^+$ & $C00006$ & $C_{21}H_{29}N_7O_{17}P_3$ \\
$ADP$ & $C00008$ & $C_{10}H_{15}N_5O_{10}P_2$ \\
Phosphoric acid & $C00009$ & $H_3PO_4$ \\
Carbon~dioxide & $C00011$ & $CO_2$ \\
Diphosphoric acid & $C00013$ & $H_4P_2O_7$ \\
Ammonia & $C00014$ & $NH_3$ \\
Pyruvate & $C00022$ & $C_3H_4O_3$ \\
Glutamate & $C00025$ & $C_5H_9NO_4$ \\
2-Oxoglutarate & $C00026$ & $C_5H_6O_5$ \\
$H^+$ & $C00080$ & $H$ \\

\hline
\end{tabular}
\label{tab2}
\end{center}
\caption{The second set of reservoirs chosen, along with their KEGG IDs \cite{Kanehisa00}. Common names are given when applicable.}
\end{table}



\subsection{The first choice of reservoirs}
\begin{figure}[h!]
\centering
\begin{subfigure}[b]{0.49\textwidth}\centering
\includegraphics[scale=0.4]{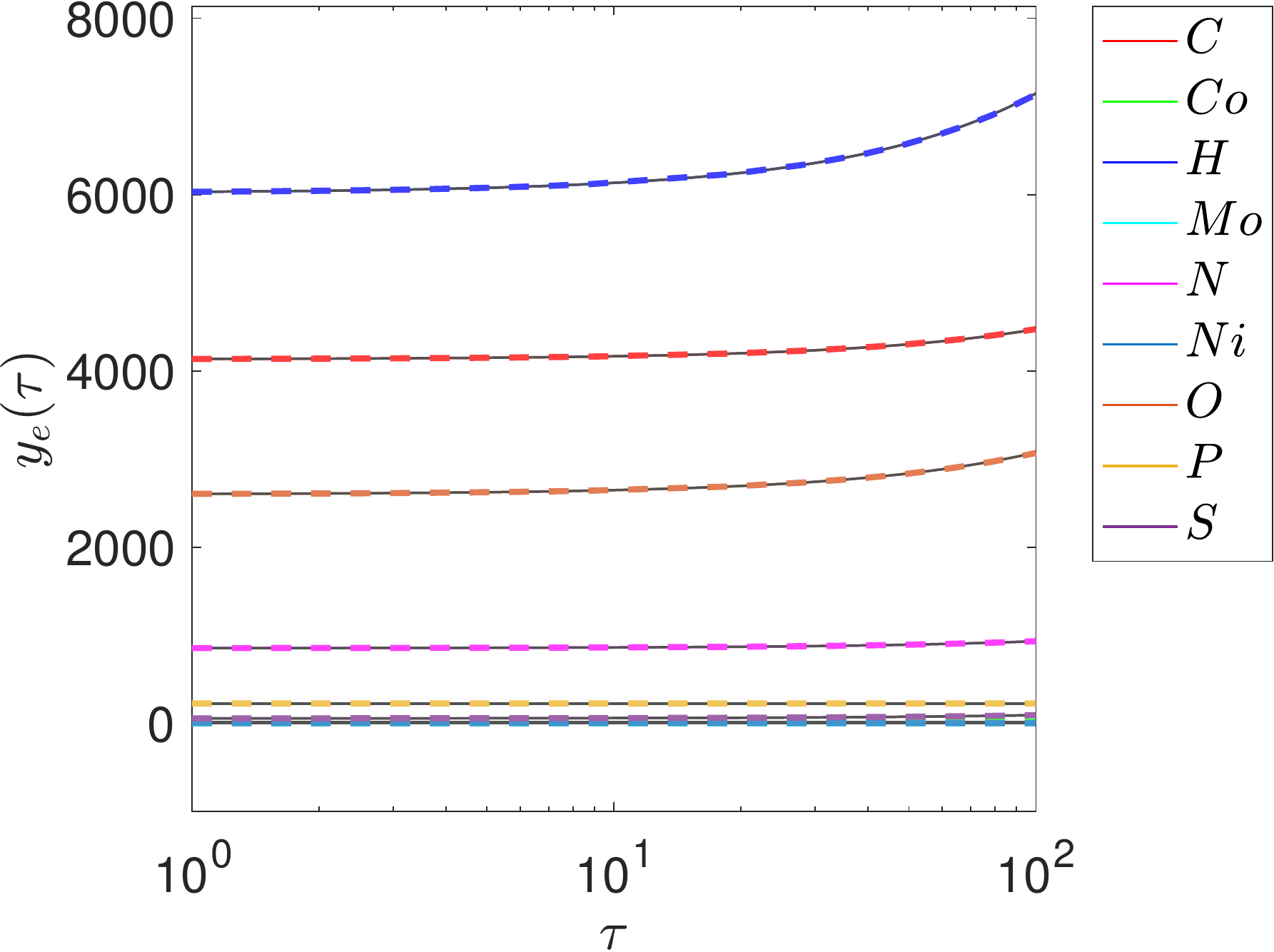}
\caption{}
\label{}
\end{subfigure}
\begin{subfigure}[b]{0.49\textwidth}\centering
\includegraphics[scale=0.4]{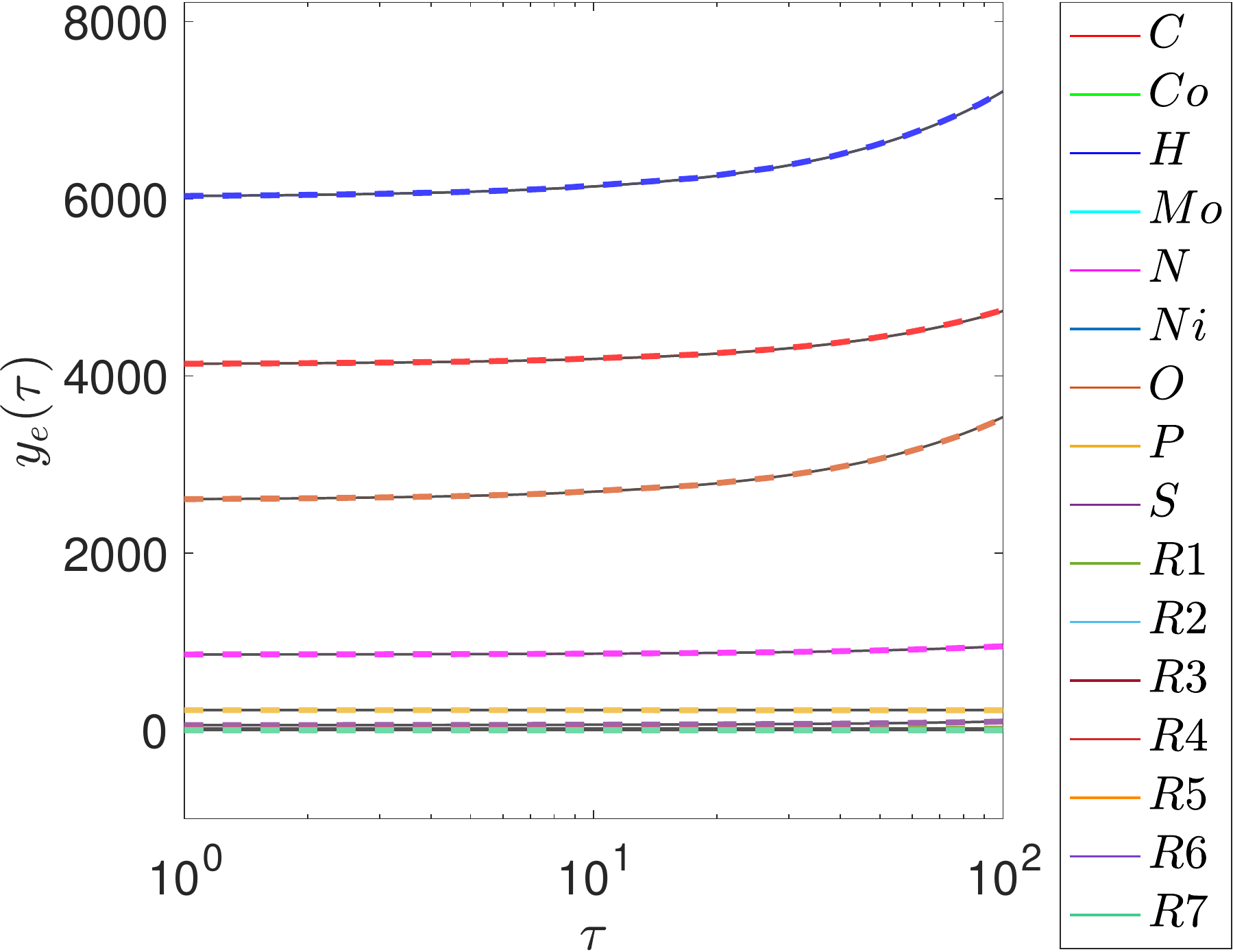}
\caption{}
\label{}
\end{subfigure}
\caption{Theoretical solutions to the slow dynamics equation (black) are compared with the numerical solutions to the rate equations (coloured), for for reservoir set 1 and the cases (a) where we take the $R$ components out of every species, and (b) where each $R_i$ is considered a conserved element.}
\label{Res1}
\end{figure}

In this case the reservoirs are limited to 8 simple compounds listed in the table \ref{tab1}. It turns out that there are 5 and 3 broken conservation laws and emergent cycles, respectively. Fig. (\ref{Res1}) show the numerical time evolutions (colored) and compared with the result from the slow dynamics equation (black), for the concentrations of elements. In (a), we remove the $R$ components from each species, while in (b) we take into account the new conserved components $R_i$. 

\subsection{The second choice of reservoirs}
\begin{figure}[h!]
\centering
\begin{subfigure}[b]{0.49\textwidth}\centering
\includegraphics[scale=0.4]{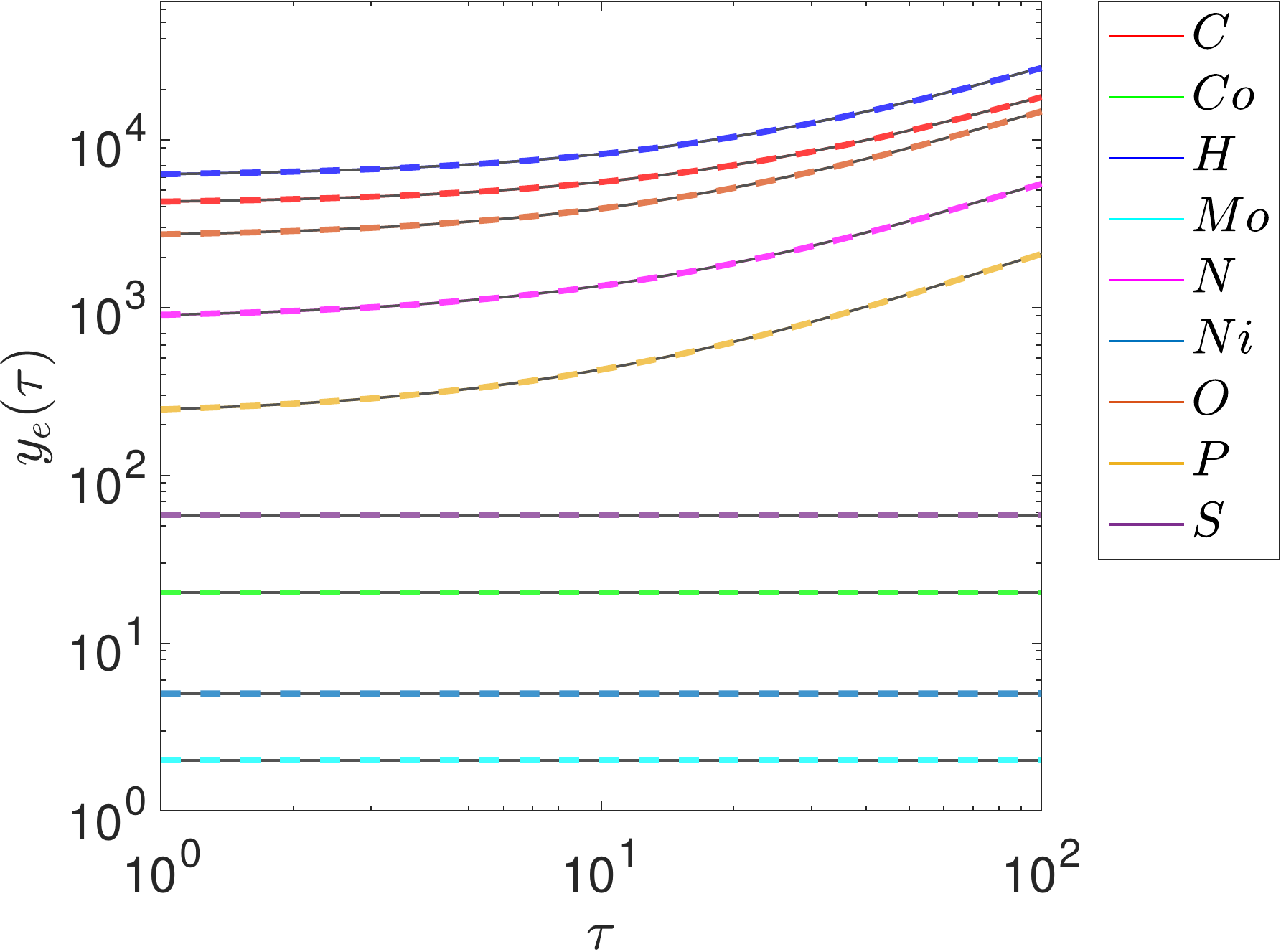}
\caption{}
\label{}
\end{subfigure}
\begin{subfigure}[b]{0.49\textwidth}\centering
\includegraphics[scale=0.4]{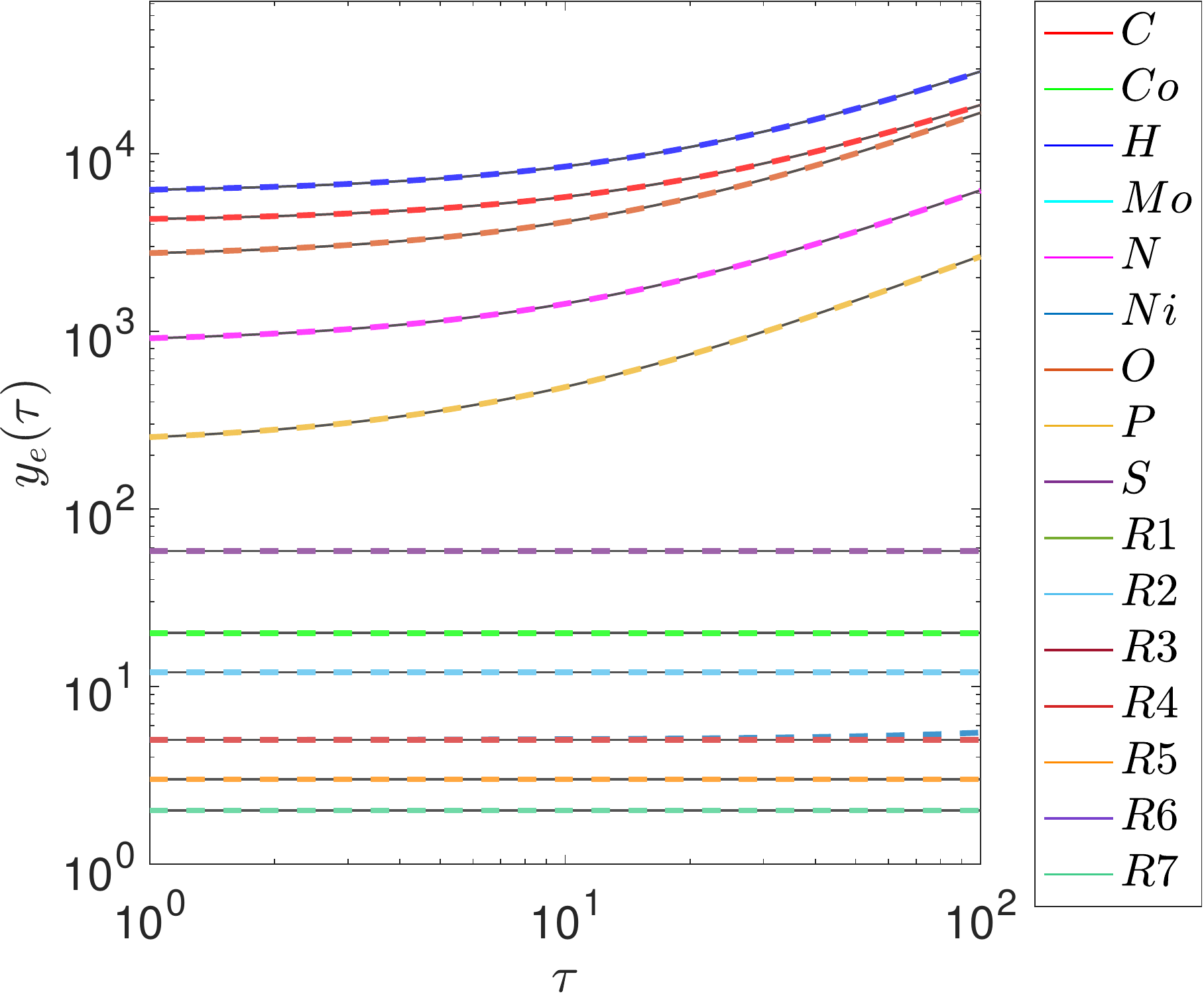}
\caption{}
\label{}
\end{subfigure}
\caption{Theoretical solutions to the slow dynamics equation (black) are compared with the numerical solutions to the rate equations (coloured), for reservoir set 2 and for the cases (a) where we take the $R$ components out of every species, and (b) where each $R_i$ is considered a conserved element.}
\label{Res2}
\end{figure}

There are 15 chosen reservoirs listed in the table \ref{tab2}. The numbers of broken conservation laws and emergent cycles turns out to be 5 and 10 respectively. The numerical time evolutions are shown in coloured bars in Fig. (\ref{Res2}) and are compared with the answer to the slow dynamics equation in black. In (a), we take each species' $R$ components out, whereas in (b), we account for the newly conserved $R_i$ components. 

\section{Connection to Circuit Theory} In the circuit theory of \cite{Avanzini23}, a CRN is reduced to a set of interconnected modules. To use it, one needs to characterize, for each module, the relationship between the flux through each emergent cycle, and the concentrations of the chemostats that force the module. Consider the circuit of Fig 3 in \cite{Avanzini23}, with internal reactions
\eq{
S + E_a & \rightleftharpoons E_a S \\
S + E_a S & \rightleftharpoons E_a S_2 \\
N_a + E_a & \rightleftharpoons E_a S ,
}
and where $N_a$ and $S$ are chemostatted. In our setup, we thus add boundary reactions
\eq{
S \rightleftharpoons S^\C \\
N_a \rightleftharpoons N_a^\C
}
From the stoichiometric point of view, one sees that there are two conserved moieties, $N_a \cong S$ and $E_a$, so in our setup there are two slow variables $\eta_S$ and $\eta_E$. Since the above reservoirs still conserve $E_a$, its density is fixed and does not couple to the single emergent cycle $S \to N_a$. For the circuit theory, one needs the flux through the emergent cycle
\eq{
\psi \equiv \epsilon J_S - \epsilon J_{N_a} 
}
where
\eq{
J_S = r_S (z_S - n^{eq}_S e^{\eta_S}/\Omega), \quad J_{N_a} = r_{N_a} (z_{N_a} - n^{eq}_{N_a} e^{\eta_S}/\Omega) ,
}
as a function of the chemostat concentrations $z_S^\C, z_{N_a}^\C$. The slow dynamics equations are \eq{
& (n^{eq}_S e^{\eta_S} + n^{eq}_{N_a} e^{\eta_S} + n^{eq}_{E_a S} e^{\eta_S+\eta_E} + 4 n^{eq}_{E_a S_2} e^{2\eta_S+\eta_E}) \p_\tau \eta_S + (n^{eq}_{E_a S} e^{\eta_S+\eta_E} + 2 n^{eq}_{E_a S_2} e^{2\eta_S+\eta_E}) \p_\tau \eta_E \notag\\
& \qquad = r_S (z_S - n^{eq}_S e^{\eta_S}/\Omega) + r_{N_a} (z_{N_a} - n^{eq}_{N_a} e^{\eta_S}/\Omega) \\
& (n^{eq}_{E_a S} e^{\eta_S+\eta_E} + 2 n^{eq}_{E_a S_2} e^{2\eta_S+\eta_E}) \p_\tau \eta_S + (n^{eq}_{E_a} e^{\eta_E} + n^{eq}_{E_a S} e^{\eta_S+\eta_E} + n^{eq}_{E_a S_2} e^{2\eta_S+\eta_E}) \p_\tau \eta_E = 0
}
In steady state we get $e^{\eta_S} = (r_S z_S+r_{N_a}z_{N_a})/(r_S n^{eq}_S/\Omega+r_{N_a} n^{eq}_{N_a}/\Omega)$ which then determines $\psi$. The resulting expression differs from that in \cite{Avanzini23} because of the boundary conditions: in their case $N_a$ and $S$ are fully chemostatted throughout the module, whereas we allow these species to dynamically evolve within the module. In their case to find $\psi$ one needs to know the bulk dynamics in the module, whereas in our case one only needs to know the boundary rates. As a result, the two methods are complementary. For simple, well-characterized modules, one can solve the dynamics as in \cite{Avanzini23}. For large, poorly-characterized modules, our technique may instead be more useful. \\


\vfill
\bibliographystyle{iopart-num}
\bibliography{../Biology}